\documentclass[12pt]{article}

\catcode`\@=11
\@addtoreset{equation}{section}

\global\arraycolsep=2pt
\usepackage{amsbsy,amssymb,latexsym,amsfonts,amsmath,cite}

\allowdisplaybreaks

\newcommand\CL{{\mathcal L}}

\newcommand\e{\mathrm{e}}

\newcommand\nn{\nonumber}

\newcommand\Phih{{\hat\Phi}}
\newcommand\Psih{{\hat\Psi}}

\newcommand{\sch}{Schr\"odinger }

\begin{document}
\begin{flushright}
\parbox{4.2cm}
{UCB-PTH-08/75 \\
OIQP-08-12 \\
NSF-KITP-08-138}
\end{flushright}

\vspace*{0.5cm}

\begin{center}
{\Large \bf 
A family of super Schr\"odinger invariant \\ 
Chern-Simons matter systems
}
\vspace*{1cm}\\
{Yu Nakayama$^{\flat,\dagger}$, Shinsei Ryu$^{\dagger}$, 
Makoto Sakaguchi$^{\sharp}$ and Kentaroh Yoshida$^{\natural}$
}
\end{center}
\vspace*{-0.2cm}
\begin{center}
$^{\flat}${\it Berkeley Center for Theoretical Physics, \\ 
University of California, Berkeley, CA 94720, USA
} 
\vspace*{0.5cm}\\
$^{\dagger}${\it Department of Physics, 
University of California, Berkeley, CA 94720, USA
} 
\vspace*{0.5cm}\\
$^{\sharp}${\it Okayama Institute for Quantum Physics\\
1-9-1 Kyoyama, Okayama 700-0015, Japan} 
\vspace*{0.5cm}\\
$^{\natural}${\it Kavli Institute for Theoretical Physics, \\ 
University of California, Santa Barbara, CA 93106, USA} 
\vspace{0.5cm}\\
{\tt nakayama@berkeley.edu} \qquad 
{\tt sryu@berkeley.edu}  \\
{\tt makoto\_sakaguchi@pref.okayama.jp} \qquad 
{\tt kyoshida@kitp.ucsb.edu} 
\end{center}

\vspace{0.8cm}

\begin{abstract} 
We investigate non-relativistic limits of the $\mathcal{N}$=3
Chern-Simons matter system in 1+2 dimensions. The relativistic theory
can generate several inequivalent super Sch\"odinger invariant theories,
depending on the degrees of freedom we choose to retain in the
non-relativistic limit. The maximally supersymmetric Schr\"odinger
invariant theory is obtained by keeping all particle degrees of
freedom. The other descendants, where particles and anti-particles
coexist, are also Schr\"odinger invariant but preserve less
supersymmetries.  Thus, we have a family of super Schr\"odinger
invariant field theories produced from the parent relativistic theory.
\end{abstract}

\thispagestyle{empty} 

\setcounter{page}{0}

\newpage

\section{Introduction} 

One of the hottest issues in AdS/CFT \cite{AdS/CFT,GKP,W} is its
application to condensed matter physics (CMP), dubbed as AdS/CMP
correspondence. Gravitational duals for condensed matter systems such as
superconductors \cite{G,HHH}, quantum hall effects \cite{KVK,DKS} and
entanglement entropy \cite{RT} are now proliferating in the recent
studies of AdS/CMP correspondence (For related progress and a review see
\cite{GP,RH,MO,replica} and \cite{review} respectively).

\medskip 

While these attempts capture some essential aspects of condensed matter
systems in a non-trivial way, a noticeable point is that the systems
dual to the gravity solutions are typically relativistic.  As a
consequence, it is not always realistic to compare them with
experimental results since condensed matter systems realized in
laboratories are typically non-relativistic. Thus, the next step in the
AdS/CFT context is to learn how to realize a non-relativistic (NR)
limit.

\medskip 

In this context, it would be important to gain much more insights into
NR CFTs \cite{Sch1,Sch2,JP,Henkel,MSW,SW,NS}.  As a break-through in
this direction, gravity duals of NR CFTs have been proposed and further
investigated in the literature
\cite{Son,BM,Gold,Barbon,Wen,HRR,MMT,ABM,HY,Schvellinger,Mazzucato,KLM,RRST}.
Among NR CFTs, theories with dynamical exponent $z=2$ are special, and
they enjoy the \sch symmetry \cite{Sch1,Sch2}, which is a NR analog of
the relativistic conformal symmetry. The name originates from the fact
that the symmetry was initially found as the maximal symmetry of a free
\sch equation. From the purely field theory viewpoint, an example of
\sch invariant field theories was proposed by Jackiw-Pi \cite{JP} based
on the NR Chern-Simons matter (CSM) system in 1+2 dimensions.

\medskip 

The \sch symmetry can accommodate supersymmetries, in which case it is
enhanced to a super \sch symmetry \cite{GGT,DH,SSch,SY1,SY2} (For super
\sch invariant gravity duals see \cite{HY}). Super \sch invariant field
theories would be important like superconformal field theories in our
attempts to embed them into string theory. Recently the index for NR
SCFTs has also been formulated \cite{Nakayama} and it plays a
significant role in classifying the theories. However, as far as we
know, the supersymmetric Jackiw-Pi model \cite{LLM}, which is obtained
by taking a NR limit of the $\mathcal{N}$=2 CSM system \cite{LLW}, is
the only example of super \sch invariant field theories with explicit
action. It would be, therefore, important to look for other examples in
order to deduce some universal features of \sch symmetry and NR
supersymmetry.

\medskip 

In this paper we consider NR limits of the $\mathcal{N}$=3
relativistic CSM system \cite{N=3}. By taking the standard NR limit with
only particle degrees of freedom, a new $\mathcal{N}$=3 super \sch
invariant CSM system is presented. By using the Noether method, the
generators of the super \sch algebra are constructed. As other
examples of NR limits, one may consider the mixed cases where particles
and anti-particles coexist. The bosonic \sch symmetry still appears but
the number of the preserved supersymmetries is reduced. This statement
seems general. The bosonic \sch symmetry is independent of the existence
of anti-particles but the supersymmetries are affected by them. We
emphasize that the way we take the non-relativistic limit affects the
symmetry of the resultant non-relativistic theory, which is quite a
novel feature.

\medskip 

Before closing the introduction, we should comment on the advantage to
begin with the $\mathcal{N}$=3 CSM system. The $\mathcal{N}$=3
supersymmetries has long been the maximum supersymmetries
in the conventional Chern-Simons matter system. This is a peculiarity to 
the (1+2)-dimensional space-time, related to an ambiguity of spin. If
we try to realize $\mathcal{N}$=3 supersymmetries in 1+3 dimensions, then 
the supersymmetries are inevitably enhanced to $\mathcal{N}$=4.
Recent developments have shown that quiver-like gauge theories admit
$\mathcal{N} =4$ or greater 
with a judicious choice of the gauge group and representations 
\cite{GW,Sangmin, ABJM}\footnote{In particular, one of the theories 
contains $Sp(2M) \times O(1) \simeq Sp(2M)$ single gauge 
field \cite{Sangmin}. We would like to thank S.~Lee for pointing out an erronous 
statement in the earlier version.}. 
Thus, as far as the conventional Chern-Simons matter systems are concerned,
it would be possible in principle to realize
all of the super \sch invariant relatives starting from the
$\mathcal{N}=3$ theory, though we will not try to make a complete list
here.

\medskip

This paper is organized as follows. Section 2 is a short review of the
relativistic $\mathcal{N}$=3 CSM system. In section 3 we study a NR
limit of the $\mathcal{N}$=3 CSM system keeping only the particles and
derive a new $\mathcal{N}$=3 super \sch invariant CSM system. In section
4 we discuss other NR limits of the $\mathcal{N}$=3 CSM system by mixing
particles and anti-particles. We still find the bosonic \sch symmetry
but the number of the preserved supersymmetries is decreased. In section
5 we discuss a consistency of NR limits in detail. Section 6 is devoted
to a summary and discussions. In Appendix A the dimensional counting is
explained. Appendix B describes the details of the spinor rotation.

\section{$\mathcal{N}=3$ relativistic CSM system} 

Our starting point is the $\mathcal{N}$=3 relativistic CSM system in 1+2
dimensions \cite{N=3}. We shall give a short review below. Our
convention used in this paper is also fixed here.

\subsection{The action of the $\mathcal{N}=3$ relativistic CSM  system}

The Lagrangian of the $\mathcal{N}$=3 relativistic CSM system \cite{N=3}
is composed of the CS term $\mathcal{L}_{\rm CS}$ and the matter part
$\mathcal{L}_{\rm M}$ as follows:\footnote{We have recovered the speed
of light $c$ explicitly. For dimensional analysis see Appendix A.}
\begin{eqnarray}
\mathcal{L}_{\rm rel} &=& \mathcal{L}_{\rm CS} + \mathcal{L}_{\rm M}\,, \nn \\ 
\mathcal{L}_{\rm CS} &=& \frac{\kappa}{4 } 
\epsilon^{\mu\nu\lambda} 
A_{\mu} F_{\nu\lambda} =
\kappa A_0 F_{12} + \frac{\kappa}{2c}\epsilon^{ij}
\partial_t A_i A_j \qquad (i,j=1,2)\,,  \nn \\ 
\mathcal{L}_{\rm M}
&=& - |D_{\mu}\phi_a |^2 
- i \bar{\psi} \gamma^{\mu}D_{\mu}\psi
- i \bar{\chi} \gamma^{\mu} D_{\mu} \chi
\nn \\
&&
-
\left(
\frac{ e^2}{ \kappa c^2} 
\right)^2 
|\phi_a |^2 
\left[
|\phi_b |^4
+ 2 v^2 
\left(|\phi_1|^2 - |\phi_2|^2 \right)
+ v^4
\right]
- \frac{ e^2}{ \kappa c^2} 
v^2
\left(
i\bar{\psi}\psi
-i\bar{\chi}\chi
\right)
\nn \\
&&
+
\frac{3e^2}{\kappa c^2} 
\left(
|\phi_1|^2 + |\phi_2|^2 
\right)
\left(
i\bar{\psi}\psi
+i\bar{\chi}\chi
\right)
-
\frac{4 i e^2}{\kappa c^2}
\left(
\phi_1 \bar{\psi}
-
\phi_2 \bar{\chi}
\right)
\left(
\phi^{\ast}_1 \psi 
-
\phi^{\ast}_2 \chi
\right)  \nn \\
&&
-
\frac{i e^2}{\kappa c^2}
\left(
\phi_1 \bar{\psi}
-
\phi_2 \bar{\chi}
\right)(-i\sigma_1)
\left(
\phi_1 \psi^{\ast} 
-
\phi_2 \chi^{\ast}
\right) \nn \\ 
&& -
\frac{i e^2}{\kappa c^2}
\left(
\phi^{\ast}_1 \bar{\psi^{\ast}}
-
\phi^{\ast}_2 \bar{\chi^{\ast}}
\right)(-i\sigma_1)
\left(
\phi^{\ast}_1 \psi
-
\phi^{\ast}_2 \chi
\right)\,.  \label{Lag}
\end{eqnarray} 
The matter action contains two complex scalar fields $\phi_a~(a=1,2)$
and two 2-component complex fermions $\psi$ and $\chi$\,. From the
quadratic parts, we identify the mass parameter $m$ as
\begin{eqnarray}
m^2c^2 \equiv 
\left(
\frac{ e^2}{ \kappa c^2} 
\right)^2 v^4\,. 
\label{mass}
\end{eqnarray}
The system is superconformal (at least at classical level) when
$v^2=0$\,.

\medskip 

In our convention, the sign of the space-time metric is $(-,+,+)$, and
we take the Dirac representation for the gamma matrices,
\begin{eqnarray}
\gamma^0=-i\sigma_3\,, \quad \gamma^1 =\sigma_1\,, \quad 
\gamma^2=\sigma_2\,. \label{Dirac}
\end{eqnarray}
The Dirac conjugate and covariant derivative are defined as 
\begin{eqnarray}
\bar{\psi} = \psi^{\dagger}\gamma^0\,, \qquad 
\bar{\psi^{\ast}} = \psi^T \gamma^0\,, \qquad 
D_i =\partial_i + \frac{ie}{c}A_i\,, \qquad 
D_0 = \frac{1}{c}\partial_t + \frac{ie}{c}A_0\,. 
\end{eqnarray}
We note that \cite{N=3} used the Majorana representation, but we have
switched to the Dirac basis that is convenient in the NR limit.  Because
of this change of spinor basis\footnote{For the details of the spinor
rotation see Appendix B.}, the additional factor $-i\sigma_1$ has
appeared in the last two lines of (\ref{Lag})\,. The $\sigma_1$ combines
with $\gamma^0=-i\sigma_3$ in the Dirac conjugate, reducing to the
standard epsilon tensor to make a Lorentz scalar out of two (complex
conjugated) fermions.

\subsection{Supersymmetries}

The relativistic $\mathcal{N}$=3 supersymmetries are given by
\begin{eqnarray}
\delta A_{\mu} &=& 
\frac{e}{\kappa c}\bar{\alpha}_1\gamma_{\mu} 
(\psi\phi_2^{\ast} - i\sigma_1 \chi^{\ast}\phi_1) 
+\frac{e}{\kappa c}
\bar{\alpha_1^{\ast}}i\sigma_1\gamma_{\mu}\left(
-i\sigma_1\psi^{\ast}\phi_2 + \chi \phi_1^{\ast}
\right) \nn \\ 
&& + \frac{e}{\kappa c}\bar{\alpha}_2\gamma_{\mu}
(-\psi\phi_1^{\ast} -i\sigma_1 \psi^{\ast}\phi_1
-\chi\phi_2^{\ast} -i\sigma_1 \chi^{\ast}\phi_2)\,, 
\label{gauge}\\ 
\delta \phi_1 &=& 
i(-\bar{\alpha_1^{\ast}} i\sigma_1 \chi 
+ \bar{\alpha}_2\psi) \,, \\ 
\delta \phi_2 &=& 
i(-\bar{\alpha}_1\psi + \bar{\alpha}_2\chi)\,, \\ 
\delta \psi &=&  -\gamma^{\mu}\alpha_1 D_{\mu}\phi_2  
+ \gamma^{\mu}\alpha_2 D_{\mu}\phi_1 \nn \\ 
&& + \frac{e^2}{\kappa c^2}(\alpha_1\phi_2-\alpha_2\phi_1)
(v^2+|\phi_1|^2-|\phi_2|^2) \nn \\ 
&& + \frac{2e^2}{\kappa c^2}(
-i\sigma_1\alpha_1^{\ast}(\phi_1)^2\phi_2^{\ast} 
+ \alpha_2\phi_1|\phi_2|^2)\,, \\ 
\delta \chi &=& -i\gamma^{\mu}\sigma_1\alpha_1^{\ast}D_{\mu}\phi_1 
+ \gamma^{\mu}\alpha_2 D_{\mu}\phi_2 \nn \\ 
&& + \frac{e^2}{\kappa c^2}
(-i\sigma_1\alpha_1^{\ast}\phi_1+\alpha_2\phi_2)
(v^2+|\phi_1|^2-|\phi_2|^2) \nn \\ 
&& + \frac{2e^2}{\kappa c^2}(i\sigma_1\alpha_1^{\ast}\phi_1^{\ast}
(\phi_2)^2 + \alpha_2|\phi_1|^2 \phi_2)\,.
\end{eqnarray}
The above supersymmetry transformation has been adjusted to our notation
from \cite{N=3}.

\medskip 

We write the 2-component complex complex spinors as 
\[
\alpha_a = \begin{pmatrix}
\alpha_a^{(1)} \\ \alpha_a^{(2)} 
\end{pmatrix}
\qquad (a=1,2)
\,, 
\]
and we impose the Majorana condition on $\alpha_2$, which relates the
first and second components as
\begin{eqnarray}
\alpha_2^{(2)} =-i(\alpha_2^{(1)})^{\ast}\, . \label{const}
\end{eqnarray}
The number of independent components for $\alpha_2$ is 2 in real
($\mathcal{N}$=1 in 1+2 dimensions). On the other hand, the number of
independent components of $\alpha_1$ is 4 in real ($\mathcal{N}$=2 in
1+2 dimensions) and corresponds to the $\mathcal{N}$=2 supersymmetries
of \cite{LLM}.  In total, the CSM system with the Lagrangian \eqref{Lag}
has $\mathcal{N}$=3 supersymmetries.

\section{NR limit of $\mathcal{N}$=3 CSM - all particle case}

Let us discuss a non-relativistic limit of the $\mathcal{N}$=3 CSM
system. The CS term is not modified via the NR limit, so we will not
touch it for a while, and we concentrate on the matter part only.

\medskip 

With the mass parameter $m$ defined in (\ref{mass}), the matter fields
are expanded as
\begin{eqnarray}
\phi_a &=&
\frac{1}{\sqrt{2m}}
\left[\,\e^{-im c^2 t} \Phi_a
+ \e^{im c^2 t} \hat{\Phi}^{\ast}_a \right] \qquad (a=1,2)\,, \nn \\
\psi &=& 
\sqrt{c}
\left[\,\e^{-i m c^2 t} \Psi
+ \e^{i m c^2 t} C \hat{\Psi}^{\ast} \right]\,, \nn \\
\chi &=& 
\sqrt{c}
\left[\,\e^{-i m c^2 t} \Upsilon
+ \e^{i m c^2 t} C \hat{\Upsilon}^{\ast}
\right]\,. \label{expand}
\end{eqnarray}
The symbol ``hat'' implies anti-particle and $C=i\sigma_2$ is a charge
conjugation matrix.

\medskip 

There are several ways to take a NR limit in accordance with the content
of the degrees of freedom held in the limit. Here we shall take a
natural choice: all of the particles are held on and all of the
anti-particles are discarded. That is,
\[
\hat{\Phi}_a=\hat{\Psi}=\hat{\Upsilon}=0\,.
\]
The truncation should be consistent and we will show that this limit is
indeed a consistent one in section \ref{c}.

\medskip 

The NR limit is obtained by substituting the decomposed fields
(\ref{expand}) into (\ref{Lag}) and then taking $c\to\infty$ limit. The
oscillating terms can be ignored and the Lagrangian is expanded in terms
of $1/c$\,. The resulting first order Lagrangian is given by
\begin{eqnarray}
\mathcal{L}_{\rm NR}
&=& i\Phi^{\ast}_1 D_t 
\Phi_1
- \frac{1}{2m} (D_i\Phi_1)^{\ast} (D_i\Phi_1) 
+ i\Phi^{\ast}_2 D_t 
\Phi_2 
- \frac{1}{2m} (D_i\Phi_2)^{\ast} (D_i\Phi_2) 
\nn \\
&& +i\Psi^{\ast}_1 D_t
\Psi_1
-\frac{1}{2m}(D_i\Psi_1)^{\ast} (D_i \Psi_1) 
+i\Upsilon^{\ast}_2 D_t 
\Upsilon_2
- \frac{1}{2m} (D_i\Upsilon_2)^{\ast}(D_i \Upsilon_2) \nn \\
&& -\frac{e}{2mc}F_{12}(|\Psi_1|^2 -|\Upsilon_2|^2)  
- \lambda \left(|\Phi_1|^4 - |\Phi_2|^4 \right) \nn \\ && 
+3 \lambda\left(|\Phi_1|^2 + |\Phi_2|^2 \right)
\left(|\Psi_1|^2 - |\Upsilon_2|^2 \right) 
-4\lambda\left(|\Phi_1|^2  |\Psi_1|^2  
-|\Phi_2|^2   |\Upsilon_2|^2 \right) \nn \\ 
&& -2i\lambda\left(\Phi_1\Phi_2\Psi_1^{\ast}\Upsilon_2^{\ast}
+ \Phi_1^{\ast}\Phi_2^{\ast}\Psi_1\Upsilon_2 \right) +\mathcal{O}(1/c^2)\,. 
\label{N=3-1}
\end{eqnarray}
Here the following quantities have been introduced 
\[
D_t \equiv cD_0 = \partial_t + ieA_0\,, \qquad 
\lambda\equiv \frac{e^2}{2m\kappa c}\,.
\]
The absolute value of the fermions depends on the ordering. We define it
as
\begin{eqnarray}
|\Psi_1|^2 \equiv \Psi_1^{\ast}\Psi_1\,, \qquad 
|\Upsilon_2|^2 \equiv \Upsilon_2^{\ast}\Upsilon_2\,. 
\end{eqnarray}

In the derivation of (\ref{N=3-1}) we used the fermion equations of
motion
\begin{eqnarray}
\Psi_2 = -\frac{1}{2mc} D_+ \Psi_1 + \mathcal{O}(1/c^2)\,, 
\qquad 
\Upsilon_1=\frac{1}{2mc} D_- \Upsilon_2 + \mathcal{O}(1/c^2)\,, 
\end{eqnarray}
and removed $\Psi_2$ and $\Upsilon_1$\,. Here we have recombined 
the spatial covariant derivatives as follows:
\begin{eqnarray}
D_{\pm} \equiv D_1 \pm i D_2\,.
\end{eqnarray} 
Hereafter we ignore the second
order and higher order corrections in terms of $1/c$\,.

\medskip 

For later purposes, we note that the Lagrangian contains an
$\mathcal{N}=2$ NR CSM theory \cite{LLM} as a subsystem by setting
\[
\Phi_1=\Upsilon_2=0\,, 
\]
up to the difference in conventions.

\subsection{Schr\"odinger symmetry}

It is turn to check symmetries of the NR Lagrangian, $\mathcal{L}_{\rm
CS}+\mathcal{L}_{\rm NR}$\,, with (\ref{N=3-1}).  We first show that the
NR Lagrangian possesses a Schr\"odinger symmetry.

\subsubsection*{The generators of \sch symmetry}

The transformation laws and corresponding charges are summarized below:
\begin{enumerate}
\item time translation - $\delta t =-a$ \\
The transformation law is 
\begin{eqnarray}
&&\delta \Phi_1 = a D_t\Phi_1\,, \quad \delta \Phi_2 = a D_t \Phi_2\,, \quad 
\delta \Psi_1 = a D_t\Psi_1\,, \quad \delta \Upsilon_2 = a D_t \Upsilon_2\,, 
\nn \\ 
&& \delta A_0=0\,, \quad \delta A_i = a c F_{0i}\,, \nn 
\end{eqnarray}
and the generator is the Hamiltonian: 
\begin{eqnarray}
&& \quad 
H = \int\!d^2x\,\biggl[\frac{1}{2m}(D_i\Phi_1)^{\ast}(D_i\Phi_1) 
+\frac{1}{2m}(D_i\Phi_2)^{\ast}(D_i\Phi_2) 
\nn \\
&& \hspace*{2.7cm}
+\frac{1}{2m}(D_i\Psi_1)^{\ast}(D_i\Psi_1) 
+\frac{1}{2m}(D_i\Upsilon_2)^{\ast}(D_i\Upsilon_2) \nn \\ 
&& \hspace*{2.7cm} 
+\frac{e}{2mc}F_{12}(|\Psi_1|^2  
-|\Upsilon_2|^2)  
+ \lambda\left(|\Phi_1|^4 - |\Phi_2|^4 \right) \nn \\ 
&& \hspace*{2.7cm} 
- 3 \lambda
\left(
|\Phi_1|^2 + |\Phi_2|^2 
\right)
\left(|\Psi_1|^2 - |\Upsilon_2|^2 
\right) 
+4\lambda
\left(|\Phi_1|^2  |\Psi_1|^2  
-|\Phi_2|^2   |\Upsilon_2|^2  
\right) \nn \\ 
&& \hspace*{2.7cm}
+2i\lambda
\left(\Phi_1\Phi_2\Psi_1^{\ast}\Upsilon_2^{\ast}
+ \Phi_1^{\ast}\Phi_2^{\ast}\Psi_1\Upsilon_2 \right)
\biggr]\,.  \label{H1}
\end{eqnarray}
\item spatial translation - $\delta x^i =a^i~(i=1,2)$ \\
The transformation law is 
\begin{eqnarray}
&& \delta \Phi_1 = -a^iD_i\Phi_1\,, \quad 
\delta \Phi_2 = -a^iD_i\Phi_2\,, \quad 
\delta \Psi_1 = -a^iD_i\Psi_1\,, \quad 
\delta \Upsilon_2 = -a^iD_i\Upsilon_2\,, 
\nn \\ 
&& \delta A_0=a^iF_{0i}\,, \quad \delta A_i = \epsilon_{ij}a^jF_{12} 
\qquad (\epsilon_{12}=-\epsilon_{21}=1)\,, \nn 
\end{eqnarray}
and the generator is the momentum: 
\begin{eqnarray} 
&& \quad P_i = \int\!d^2x\, p_i\,, \nn \\
&& \quad \quad p_i \equiv  -\frac{i}{2}\bigl[
\Phi_1^{\ast}D_i\Phi_1 -(D_i\Phi_1)^{\ast}\Phi_1 
+ \Phi_2^{\ast}D_i\Phi_2 - (D_i\Phi_2)^{\ast}\Phi_2 \nn \\ 
&& \qquad \quad\qquad  + \Psi_1^{\ast}D_i\Psi_1 - (D_i\Psi_1)^{\ast}\Psi_1 
+ \Upsilon_2^{\ast}D_i\Upsilon_2 - (D_i\Upsilon_2)^{\ast}\Upsilon_2
\bigr]\,.
\end{eqnarray}
\item spatial rotation - $\delta x^i = \theta \epsilon^{ij}x^j$ \\
The transformation law is 
\begin{eqnarray}
&& \delta \Phi_1 = -\theta \epsilon_{ij}x^iD_j\Phi_1\,, \quad 
\delta \Phi_2 = -\theta \epsilon_{ij}x^iD_j\Phi_2\,, \nn \\ 
&& \delta \Psi_1 = -\theta \epsilon_{ij}x^iD_j\Psi_1 
-\frac{i}{2}\theta\Psi_1\,, \quad 
\delta \Upsilon_2 = -\theta \epsilon_{ij}x^iD_j\Upsilon_2 
+\frac{i}{2}\theta\Upsilon_2\,, \nn \\
&& \delta A_0 =\theta \epsilon_{ij}x^i F_{0j}\,, \quad 
\delta A_i =\theta x^i F_{12}\,, \nn 
\end{eqnarray}
and the generator is the angular momentum: 
\begin{eqnarray}
&& \quad J = \int\!d^2x\,\left[
\epsilon^{ij}x_ip_j + \frac{1}{2} (|\Psi_1|^2 - |\Upsilon_2|^2)
\right]\,.  \label{J}
\end{eqnarray}
Note that for the fermionic fields the spin operators are contained.
The relative sign of the spin part are fixed from the last term in the
Lagrangian (\ref{N=3-1}).
\item Galilean boost - $\delta x^i = v^i t$  \\ 
The transformation law is 
\begin{eqnarray}
&& \delta \Phi_1 = (imv^ix^i-tv^iD_i)\Phi_1\,, \quad 
\delta \Phi_2 = (imv^ix^i-tv^iD_i)\Phi_2\,, \nn \\ 
&&  \delta \Psi_1 = (imv^ix^i-tv^iD_i)\Psi_1\,, \quad 
\delta \Upsilon_2 = (imv^ix^i-tv^iD_i)\Upsilon_2\,, \nn \\ 
&& \delta A_0 = t v^i F_{0i}\,, \quad 
\delta A_i = t \epsilon_{ij} v^j F_{12}\,, \nn 
\end{eqnarray}
and the corresponding generator is 
\begin{eqnarray}
&& \quad G_i = t P_i - m \int\!d^2x\, x_i \rho\,. 
\end{eqnarray}
\item dilatation - $\delta t = 2a t\,,~\delta x^i = a x^i$ \\ 
The transformation law is 
\begin{eqnarray}
&& \delta \Phi_1 = - a \left[1+x^iD_i+2t D_t\right]\Phi_1\,, \quad 
\delta \Phi_2 = - a \left[1+x^iD_i+2t D_t\right]\Phi_2\,, \nn \\ 
&& \delta \Psi_1 = - a \left[1+x^iD_i+2t D_t\right]\Psi_1\,, \quad 
\delta \Upsilon_2 = - a \left[1+x^iD_i+2t D_t\right]\Upsilon_2\,, \nn \\
&& \delta A_0 = a x^i F_{0i}\,, \quad 
\delta A_i = a (\epsilon_{ij}x^jF_{12}-2ct F_{0i})\,. \nn 
\end{eqnarray}
The corresponding generator is
\begin{eqnarray}
&& \quad D = - 2tH + \int\!d^2x\,x^ip_i\,.
\end{eqnarray}
\item special conformal transformation 
- $\delta t = a t^2\,,~\delta x^i= a t x^i$ \\ 
The transformation law is  
\begin{eqnarray}
&& \delta \Phi_1 = -\left(at - \frac{i}{2}ma (x^i)^2 +atx^iD_i +at^2 D_t
\right)\Phi_1\,, \nn \\ 
&& \delta \Phi_2 = -\left(at - \frac{i}{2}ma (x^i)^2 +atx^iD_i +at^2 D_t
\right)\Phi_2\,, \nn \\ 
&& \delta \Psi_1 = -\left(at - \frac{i}{2}ma (x^i)^2 +atx^iD_i +at^2 D_t
\right)\Psi_1\,, \nn \\ 
&& \delta \Upsilon_2 = -\left(at - \frac{i}{2}ma (x^i)^2 +atx^iD_i +at^2 D_t
\right)\Upsilon_2\,, \nn \\ 
&& \delta A_0 = atx^iF_{0i}\,, \quad 
\delta A_i = at \epsilon_{ij}x^jF_{12} - at^2cF_{0i}\,, \nn 
\end{eqnarray}
and the corresponding generator is 
\begin{eqnarray}
&& \quad 
K = t^2 H + tD - \frac{1}{2}m\int\!d^2x\,(x^i)^2\rho\,,
\end{eqnarray} 
where $\rho$ is a particle density and defined as 
\begin{eqnarray}
 \rho \equiv |\Phi_1|^2 + |\Phi_2|^2 + |\Psi_1|^2 + |\Upsilon_2|^2\,.
\end{eqnarray}
Note that it carries no space-time index in comparison with the
relativistic conformal case.
\end{enumerate}

\subsubsection*{$U(1)$ symmetries}

In addition to the above generators, the mass operator 
\begin{eqnarray}
 M = m \int\!d^2x\,\rho 
\end{eqnarray}
is also a conserved quantity as a part of the Galilean algebra (or more
precisely Bargmann algebra). Here $\rho$ is the number density and its
integral is the total number. Hence $M$ just gives the total mass.

\medskip 

The conservation of $M$ is related to $U(1)$ symmetries. In this case
the following four $U(1)$s may be found:
\vspace*{0.5cm} 
\[
\begin{array}{c|cccc}
\hline
 & \Phi_1 & \Phi_2 & \Psi_1 & \Upsilon_2 \\ 
\hline  
U(1)_1 & 0 & 1 & 1 & 0 \\ 
U(1)_2 & 1 & 0 & 0 & 1 \\ 
U(1)_3 & 1 & 0 & 1 & 0 \\ 
U(1)_4 & 0 & 1 & 0 & 1 \\  
\hline 
\end{array}
\vspace*{0.5cm}
\]
Thus, we have the four conserved quantities:
\begin{eqnarray}
&& I_1 = \int\!d^2x\,\left(|\Phi_2|^2+|\Psi_1|^2\right)\,, \quad 
I_2 = \int\!d^2x\,\left(|\Phi_1|^2+|\Upsilon_2|^2\right)\,, \nn \\ 
&& I_3 = \int\!d^2x\,\left(|\Phi_1|^2+|\Psi_1|^2\right)\,, \quad 
I_4 = \int\!d^2x\,\left(|\Phi_2|^2+|\Upsilon_2|^2\right)\,. \label{u1}
\end{eqnarray}
The conservation of $M$ follows from $I_1 + I_2$\,. 

\medskip 

Note that one of $U(1)$'s can be absorbed by using the simultaneous
phase transformation (i.e. $I_1+I_2=I_3+I_4$), so there remains
$U(1)^3$\,. By taking a linear combination of (\ref{u1}), we can fix the
three $U(1)$'s. One of them should be the mass parameter $M$\,. We
choose the remaining two as follows:
\begin{eqnarray}
&& N_{\rm B} 
= \int\!d^2x\,\left(|\Phi_2|^2 - |\Phi_1|^2\right)\,,  \\ 
&& N_{\rm F} 
= \int\!d^2x\,\left(|\Psi_1|^2 - |\Upsilon_2|^2\right)\,. 
\end{eqnarray}
We have seen that $N_{\rm F}$ generates the spin from the expression
(\ref{J})\,. We will see that supercharges are eigenstates of $N_{\rm
B}$ and $N_{\rm F}$\,.

\subsubsection*{The Poisson brackets}

In order to study the algebra, we compute the Poisson brackets of the
above generators. For the matter fields, by using the classical Poisson
brackets
\begin{eqnarray}
&& \{\Phi_1(x),\Phi_1^{\ast}(x')\}_{\rm PB} 
= - \{\Phi_1^{\ast}(x),\Phi_1(x')\}_{\rm PB} 
= -i\delta^{(2)}(x-x')\,, \nn \\ 
&& \{\Phi_2(x),\Phi_2^{\ast}(x')\}_{\rm PB} 
= - \{\Phi_2^{\ast}(x),\Phi_2(x')\}_{\rm PB}  
= -i\delta^{(2)}(x-x')\,, \nn \\ 
&& \{\Psi_1(x),\Psi_1^{\ast}(x')\}_{\rm PB} 
= \{\Psi_1^{\ast}(x),\Psi_1(x')\}_{\rm PB} 
= -i\delta^{(2)}(x-x')\,, \nn \\
&& \{\Upsilon_2(x),\Upsilon_2^{\ast}(x')\}_{\rm PB} 
= \{\Upsilon_2^{\ast}(x),\Upsilon_2(x')\}_{\rm PB}  
= -i\delta^{(2)}(x-x')\,, \nn
\end{eqnarray}
one can compute the Poisson brackets of the bosonic generators 
\begin{eqnarray}
H,~P_i,~J,~G_i,~D,~K,~M,~N_{\rm B},~N_{\rm F}\,.
\end{eqnarray}
The number of the generators is $11~(=1+2+1+2+1+1+1+1+1)$\,. Henceforth
we represent the Poisson bracket for bosonic (fermionic) generators by
$[~,~]$ ($\{~,~\}$).

\medskip 

The treatment of the gauge field is more involved. Note that $A_i$
appears in the generators but $A_0$ does not. For $A_i$\,, by solving
the equation of motion for $A_0$ (the Gauss law constraint)\,,
\[
F_{12} = \frac{e}{\kappa}\rho\,,
\]
we can obtain the explicit expression,
\begin{eqnarray}
&& A_i (t,x^i) = -\frac{e}{\kappa}\epsilon_{ij}\partial_j\int\!d^2y\,G(x-y)
\rho(t,y^i)\,, 
\label{Ai} \\  
&& \qquad G(x-y) = \frac{1}{2\pi}\ln|x-y|\,, \quad 
\partial_i^2G(x-y) = \delta^{(2)}(x-y)\,. 
\end{eqnarray}
Using (\ref{Ai}) we can compute the Poisson bracket including $A_i$\,.

\subsubsection*{The \sch algebra}

The resulting algebra is 
\begin{eqnarray}
&& [P_i,P_j]=[P_i,H]=[J,H]=[G_i,G_j]=0\,, \nn \\ 
&& [J,P_i] = \epsilon_{ij}P_j\,, \quad 
[J,G_i] = \epsilon_{ij}G_j\,, \quad 
[P_i,G_j] = \delta_{ij} M\,, \quad 
[G_i,H] = - P_i\,, \nn \\ 
&& [D,H] = 2H\,, \quad [D,K] = -2K\,, \quad [K,H] = - D\,, \nn \\
&&  [D,P_i] = P_i\,, \quad [D,J]=0\,, \quad 
[D,G_i] = - G_i\,, \nn \\
&& [K,P_i] = G_i\,, \quad [K,J]=[K,G_i]=0\,, \quad 
[M,\ast]=[N_{\rm B},\ast]=[N_{\rm F},\ast]=0\,, 
\label{comm}
\end{eqnarray}
where the symbol $\ast$ in the Poisson brackets imply any bosonic
generators. This is the Schr\"odinger algebra. The three Poisson
brackets in the third line of (\ref{comm}) describe the algebra of the
one-dimensional conformal group $SO(2,1)$\,. The algebra (\ref{comm})
contains the Bargmann algebra spanned by $\{H,P_i,J,G_i,M\}$ as a
subalgebra. This is a central extension of the Galilean algebra spanned
by $\{H,P_i,J,G_i\}$ with the mass generator $M$\,.

\medskip 

It is easy to check the conservation of the generators with (\ref{comm})
and the following Hamilton equation:
\begin{eqnarray}
\hspace*{2cm}\frac{d A}{dt} = \frac{\partial A}{\partial t} + [A,H] \qquad 
(A:~\mbox{any~generator})\,. \label{conserved}
\end{eqnarray}
Now that $G_i$, $D$ and $K$ explicitly depend on the time $t$\,, they do
not commute with the Hamiltonian $H$ but are still conserved.

\subsection{Supersymmetries}

Let us consider a non-relativistic limit of the original relativistic
supersymmetries. In the non-relativistic limit, the relativistic
transformation law can be expanded in terms of $c$\,. The
non-relativistic analog of supersymmetry transformations are determined
order by order. For this purpose it is helpful to recall that
\[
v^2 = \frac{\kappa c^3 m}{e^2} \quad  \mbox{and}
\quad A_0 \sim \mathcal{O}(1/c)\,.
\] 

\subsubsection*{The NR supersymmetries}

The supersymmetry transformation at the leading order is given by
\begin{eqnarray}
\delta_1 \Phi_1 &=& \sqrt{2mc}\, (-i\alpha_1^{(1)} \Upsilon_2 
+ \alpha_2^{(1)\ast} \Psi_1)\,, \nn \\ 
\delta_1 \Phi_2 &=& -\sqrt{2mc}\,
(\alpha_1^{(1)\ast}\Psi_1+\alpha_2^{(2)\ast}\Upsilon_2)\,, 
\nn \\
\delta_1 \Psi_1 &=& \sqrt{2mc}\,(\alpha_1^{(1)}\Phi_2
-\alpha_2^{(1)}\Phi_1)\,, \nn \\
\delta_1 \Upsilon_2 &=& \sqrt{2mc}\,(-i\alpha_1^{(1)\ast} \Phi_1
+ \alpha_2^{(2)}\Phi_2)\,, \nn \\ 
\delta_1 A_0 &=& \frac{e}{\sqrt{2mc}\,\kappa}\,
\Bigl[ \alpha_1^{(1)\ast}\Psi_1\Phi_2^{\ast} 
-i \alpha_1^{(1)\ast}\Upsilon_2^{\ast}\Phi_1 
- \alpha_1^{(1)}\Psi_1^{\ast}\Phi_2 
-i \alpha_1^{(1)}\Upsilon_2\Phi_1^{\ast} \nn \\ 
&& \qquad \qquad  -\alpha_2^{(1)\ast}\Psi_1\Phi_1^{\ast} 
-i \alpha_2^{(2)\ast}\Psi_1^{\ast}\Phi_1 
-\alpha_2^{(2)\ast}\Upsilon_2\Phi_2^{\ast} 
-i \alpha_2^{(1)\ast}\Upsilon_2^{\ast}\Phi_2 
\Bigr]\,, \nn \\
\delta_1 A_i &=& 0\,. \nn  
\end{eqnarray}
Note that the reality of $A_0$ is preserved under the condition
(\ref{const})\,. The leading supersymmetry is often called the
kinematical supersymmetry.

\medskip

The second supersymmetry transformation is obtained from the
next-to-leading order in the NR limit and given by
\begin{eqnarray}
\delta_2 \Phi_1 &=&  \frac{1}{\sqrt{2mc}}
(\alpha_2^{(2)\ast}D_+\Psi_1 + i\alpha_1^{(2)} D_-\Upsilon_2)\,,
\nn \\ 
\delta_2 \Phi_2 &=& \frac{1}{\sqrt{2mc}}
(-\alpha_1^{(2)\ast}D_+\Psi_1 
+ \alpha_2^{(1)\ast}D_-\Upsilon_2 )\,,  \nn \\
\delta_2 \Psi_1 &=& -\frac{1}{\sqrt{2mc}}(
\alpha_1^{(2)}D_-\Phi_2 - \alpha_2^{(2)}D_-\Phi_1)\,, \nn \\
\delta_2 \Upsilon_2 &=& \frac{1}{\sqrt{2mc}}(
-i\alpha_1^{(2)\ast}D_+\Phi_1 + \alpha_2^{(1)}D_+\Phi_2
)\,,
\nn \\ 
\delta_2 A_0 &=& \frac{e}{(2mc)^{3/2}\kappa}\Bigl[
-\alpha_1^{(2)\ast}(D_+\Psi_1)\Phi_2^{\ast} 
-i\alpha_1^{(2)\ast}(D_-\Upsilon_2)^{\ast}\Phi_1 \nn \\ 
&& \hspace*{2.1cm} +\alpha_1^{(2)}(D_+\Psi_1)^{\ast}\Phi_2 
-i\alpha_1^{(2)}(D_-\Upsilon_2)\Phi_1^{\ast} \nn \\ 
&& \hspace*{2.1cm}
+\alpha_2^{(2)\ast}(D_+\Psi_1)\Phi_1^{\ast} 
+i \alpha_2^{(1)\ast}(D_+\Psi_1)^{\ast}\Phi_1 \nn \\ 
&& \hspace*{2.1cm}
- \alpha_2^{(1)\ast}(D_-\Upsilon_2)\Phi_2^{\ast} 
-i \alpha_2^{(2)\ast}(D_-\Upsilon_2)^{\ast}\Phi_2
\Bigr]\,, \nn \\
\delta_2 A_+ &=& 
\frac{2ie}{\sqrt{2mc}\,\kappa} 
\Bigl[
\alpha_1^{(2)}\Psi_1^{\ast}\Phi_2 
+i\alpha_1^{(2)}\Upsilon_2 \Phi_1^{\ast} 
+i\alpha_2^{(1)\ast}\Psi_1^{\ast}\Phi_1  
+\alpha_2^{(1)\ast}\Upsilon_2\Phi_2^{\ast} 
\Bigr]\,, 
\nn \\ 
\delta_2 A_- &=& 
-\frac{2ie}{\sqrt{2mc}\,\kappa} 
\Bigl[
-\alpha_1^{(2)\ast}\Psi_1\Phi_2^{\ast} 
+i\alpha_1^{(2)\ast}\Upsilon_2^{\ast}\Phi_1 
+\alpha_2^{(2)\ast}\Psi_1\Phi_1^{\ast} 
+i \alpha_2^{(2)\ast}\Upsilon_2^{\ast}\Phi_2
\Bigr]\,.  \nn  
\end{eqnarray} 
The next-to-leading supersymmetry is often called the dynamical
supersymmetry.

\medskip 
 
Here we should notice that $\alpha_1^{(1)}$ and $\alpha_1^{(2)}$ are
completely separated as in the case of \cite{LLM} (actually those
correspond to $\mathcal{N}$=2 of \cite{LLM}), but $\alpha_2^{(1)}$
(equivalently $\alpha_2^{(2)}$) appears in both the leading and the
next-to-leading supersymmetries. One might naively expect that the NR
supersymmetry should be enhanced after taking the NR limit. It is not
the case, however. By directly checking the symmetry, we can realize
that the leading supersymmetry is preserved while the next-to-leading
one (for $\alpha_2$) is broken due to the presence of the interaction
potential. Indeed, the last potential term in (\ref{N=3-1}) breaks the
symmetry\footnote{In the free field theory limit, the next-to-leading
supersymmetry is a symmetry, but the algebra does not close as a
conventional supersymmetry algebra.}.

\medskip

\subsubsection*{Supercharges}

By using the Noether method, we can construct supercharges corresponding
to the above supersymmetry transformations.

\medskip 

The supercharges for the two leading supersymmetries are 
\begin{eqnarray}
&& Q_1^{(1)} = \sqrt{2m}\int\!d^2x\,\left[
\Phi_1^{\ast}\Upsilon_2 -i\Phi_2\Psi_1^{\ast}
\right]\,, \\ 
&& Q_1^{(2)} = \sqrt{2m}\int\!d^2x\,\left[
\Phi_2^{\ast}\Upsilon_2 +i\Phi_1\Psi_1^{\ast}
\right]\,,
\end{eqnarray}
(up to rescaling $\sqrt{c}$) and the supersymmetry transformations for
the matter fields are generated by, respectively,
\[
\alpha_1^{(1)}Q_1^{(1)} + Q_1^{(1)\ast}\alpha_1^{(1)\ast}\,, \qquad 
\alpha_2^{(1)}Q_1^{(2)} + Q_1^{(2)\ast}\alpha_2^{(1)\ast}\,.
\]
For example, the transformation in terms of the first charge
$\delta_1^{(1)}$ for $\Phi_1$ is given by
\[
\delta_1^{(1)}\Phi_1 = [\Phi_1,\alpha_1^{(1)}Q_1^{(1)} 
+ Q_1^{(1)\ast}\alpha_1^{(1)\ast}]\,.
\] 
The next-to-leading supercharge is given by 
\begin{eqnarray}
Q_2 = \frac{1}{\sqrt{2m}}\int\!d^2x\,
\left[
-\Phi_1^{\ast}D_-\Upsilon_2 -i\Phi_2(D_+\Psi_1)^{\ast}
\right]\,,
\end{eqnarray}
(up to rescaling $1/\sqrt{c}$) and the transformation for the matter
fields are generated by
\[
\alpha_1^{(2)}Q_2 + Q_2^{\ast}\alpha_1^{(2)\ast}\,.
\]

\subsubsection*{The algebra with supercharges}

The Poisson brackets including supercharges only are 
\begin{eqnarray}
&& \{Q_1^{(1)},Q_1^{(1)\ast}\} = \{Q_1^{(2)},Q_1^{(2)\ast}\} = -2iM\,, 
\quad \{Q_1^{(1)},Q_1^{(2)}\} = \{Q_1^{(1)},Q_1^{(2)\ast}\} =0\,, \nn \\
&& \{Q_1^{(1)},Q_2\} =0\,, \quad \{Q_1^{(1)},Q_2^{\ast}\}=P_+\,, \quad 
\{Q_1^{(2)},Q_2\} = \{Q_1^{(2)},Q_2^{\ast}\}=0 \,, \nn \\ 
&& \{Q_2,Q_2^{\ast}\} = -iH\,. 
\end{eqnarray} 
In the derivation of the last Poisson bracket, we have used the Gauss
law constraint
\begin{eqnarray}
F_{12} = \frac{e}{\kappa}\left(|\Phi_1|^2 + |\Phi_2|^2 
+ |\Psi_1|^2 + |\Upsilon_2|^2\right)\,. \label{Gauss}
\end{eqnarray}

\medskip 

The Poisson brackets of the bosonic generators ($H,P_i,J,G_i,M,N_{\rm
B},N_{\rm F}$) and the supercharges are
\begin{eqnarray}
&& [P_i,Q_1^{(1)}] = [P_i,Q_1^{(2)}] = [P_i,Q_2] = 0\,, \quad 
[H,Q_1^{(1)}] = [H,Q_1^{(2)}] = [H,Q_2] = 0\,, \nn \\ 
&& [Q_1^{(1)},J] =\frac{i}{2} Q_1^{(1)}\,, \quad 
[Q_1^{(2)},J] =\frac{i}{2} Q_1^{(2)}\,, \quad 
[J,Q_2] =\frac{i}{2} Q_2\,, \nn \\ 
&& [G_i,Q_1^{(1)}] = [G_i,Q_1^{(2)}] = 0\,, \quad 
[Q_2,G_+] = -iQ_1^{(1)}\,, \quad [Q_2,G_-] =0\,, \nn  \\ 
&& [N_{\rm B},Q_1^{(1)}]=iQ_1^{(1)}\,, \quad 
[N_{\rm F},Q_1^{(1)}]=-iQ_1^{(1)}\,, 
\quad [M,Q_1^{(1)}]=0\,, \nn \\  
&& [N_{\rm B},Q_1^{(2)}]=-iQ_1^{(2)}\,, \quad 
[N_{\rm F},Q_1^{(2)}]=-iQ_1^{(2)}\,, 
\quad [M,Q_1^{(2)}]=0\,, \nn \\ 
&& [N_{\rm B},Q_2]=iQ_2\,, \quad [N_{\rm F},Q_2]=-iQ_2\,, \quad 
[M,Q_2]=0\,.  
\end{eqnarray}
The above algebra and the Bargmann algebra give a closed subalgebra
called super Bargmann algebra. It is worthwhile mentioning that
supercharges are eigenstates of $N_{\rm B}$ and $N_{\rm F}$\,. Thus
$N_{\rm B}$ and $N_{\rm F}$ can be interpreted as R-charges.

\subsubsection*{Superconformal symmetry} 

Now there are conformal generators $D$ and $K$\,. The following Poisson
brackets yield a closed algebra:
\begin{eqnarray}
[D,Q_1^{(1)}]=[K,Q_1^{(1)}]=[D,Q_1^{(2)}]=[K,Q_1^{(2)}]=0\,, \quad 
[D,Q_2] = Q_2\,,
\end{eqnarray}
but, for the Poisson bracket of $K$ and $Q_2$\,, we have to introduce a
new generator $S$ describing superconformal symmetry as follows:
\begin{eqnarray}
[K, Q_2] =-iS\,. 
\end{eqnarray}
Here $S$ is explicitly given by 
\begin{eqnarray} 
S = it Q_2 - \sqrt{\frac{m}{2}}\int\!d^2x\,x^- \left(\Phi_1^{\ast}\Upsilon_2 
-i\Phi_2\Psi_1^{\ast}\right)\,,
\end{eqnarray}
where $x^- = x^1 -ix^2$\,. 

\medskip 

While the superconformal transformation for the matter fields is
generated by $\beta S + S^{\ast}\beta^{\ast}$ with the Poisson bracket
as $\delta \Phi = [\Phi, \beta S + S^{\ast}\beta^{\ast}]$\,, it is not
so trivial to fix the gauge field transformation and it has not been
done even for the $\mathcal{N}$=2 case \cite{LLM}.

\medskip 

A key observation is to notice that the explicit $t$-dependence in front
of $Q_2$ in the superconformal charge $S$ gives the additional terms in
the superconformal variation in the action, but those are canceled out
by the second term contribution in $S$\,. This cancellation mechanism
works also for the gauge field transformation and this strategy enables
us to derive the additional transformation explicitly.

\medskip 

The superconformal transformation law is shown to be
\begin{eqnarray}
\delta \Phi_1&=&
-\frac{1}{\sqrt{2m}}\,t\beta D_-\Upsilon_2
+i\sqrt{\frac{m}{2}}\,x^-\beta\Upsilon_2\,, \nn \\
\delta \Phi_2&=&
\frac{1}{\sqrt{2m}}\,it\beta^* D_+\Psi_1
+\sqrt{\frac{m}{2}}\,x^+\beta^* \Psi_1\,, \nn \\
\delta \Psi_1&=&
-\frac{1}{\sqrt{2m}}\,it\beta D_-\Phi_2
-\sqrt{\frac{m}{2}}\,x^- \beta \Phi_2\,, \nn \\
\delta \Upsilon_2&=&
-\frac{1}{\sqrt{2m}}\,t\beta^* D_+\Phi_1
+i\sqrt{\frac{m}{2}}\,x^+ \beta^* \Phi_1\,, \nn \\
\delta A_0&=&
\frac{e}{(2m)^{3/2}\kappa c}
\Big[
it\beta^* D_+\Psi_1\Phi_2^*
-t\beta^*(D_-\Upsilon_2)^*\Phi_1 \nn \\ 
&& \hspace*{2cm} +it\beta (D_+\Psi_1)^*\Phi_2
+t\beta D_-\Upsilon_2 \Phi_1^* \cr 
&& \hspace*{2cm} 
+mx^-\beta(\Phi_2\Psi_1^* +i\Phi_1^*\Upsilon_2)
-mx^+\beta^*(\Phi_2^*\Psi_1 -i\Phi_1\Upsilon_2^*)
\Big]\,, \nn \\
\delta A_+&=&\frac{2ie}{\sqrt{2m}\,\kappa}\Big[
it\beta\Psi_1^*\Phi_2 
-t\beta\Upsilon_2\Phi_1^*
\Big]\,, \nn \\
\delta A_-&=&-\frac{2ie}{\sqrt{2m}\,\kappa}\Big[
it\beta^*\Psi_1\Phi_2^* 
+t\beta^*\Upsilon_2^*\Phi_1
\Big]\,. \label{sc}
\end{eqnarray} 
It is easy to check that the action is indeed invariant under this
transformation (\ref{sc})\,. As a side remark, we note that the
transformation law above (up on a trivial truncation of fields) gives
the missing gauge field transformations in the $\mathcal{N} =2$ NR CSM
system \cite{LLM} under the superconformal transformation.

\medskip 

The Poisson brackets of $S$ and the bosonic generators are 
\begin{eqnarray}
&& [S,H] = -iQ_2\,, \quad [P_+,S] = Q_1^{(1)}\,, \quad 
[P_-,S] = 0\,, \quad [J,S] =\frac{i}{2} S\,,  \nn \\ 
&&  [S,G_i] = 0\,, \quad [S,D] = S\,, \quad [S,K] = 0\,, \nn \\  
&& [N_{\rm B},S]=iS\,, \quad [N_{\rm F},S]=-iS\,, \quad 
[M,S]=0\,. 
\end{eqnarray}
The first Poisson bracket indicates the conservation of $S$ as in
(\ref{conserved})\,.

\medskip 

Similarly, the Poisson brackets with the supercharges are found to be 
\begin{eqnarray}
&& \{S,Q_{1}^{(1)}\} = \{S,Q_1^{(2)}\} = \{S,Q_2\} 
=\{S,Q_1^{(2)\ast}\} = 0\,, \nn \\ 
&& \{S,S^{\ast}\} = iK\,, \quad \{S,Q_1^{(1)\ast}\} = -i G_- \,, \quad \nn \\ 
&& \{S,Q_2^{\ast}\} 
= \frac{i}{2}\left[iD-J + N_{\rm B} - \frac{1}{2}N_{\rm F}\right]\,. 
\end{eqnarray}
Thus, we have shown that a set of the generators 
\[
H,~P_i,~J,~G_i,~D,~K,~M,~N_{\rm B},~N_{\rm F},~Q_1^{(1)},~ Q_1^{(2)},~Q_2,~S
\]
spans a super Schr\"odinger algebra with 8 supercharges (in real
components). We would like to emphasize this super \sch algebra has not
appeared before in the literature and it is a novel algebra.

\subsection{The positivity of the Hamiltonian}

It is valuable to see the positivity of the Hamiltonian (\ref{H1}).
From the expression of (\ref{H1}), the positivity is quite non-trivial
since (\ref{H1}) contains a quartic potential with the negative
sign. Nevertheless, the Hamiltonian is positive definite as required
from the supersymmetry.

\medskip 

With the Gauss law constraint (\ref{Gauss}), we can drastically simplify
(\ref{H1}) to
\begin{eqnarray}
H &=& \int\!d^2x\,\biggl[\frac{1}{2m}(D_+\Phi_1)^{\ast}D_+\Phi_1 
+ \frac{1}{2m}(D_-\Phi_2)^{\ast}D_-\Phi_2 \nn \\ 
&& \hspace*{1cm} + \frac{1}{2m}(D_-\Psi_1)^{\ast}D_-\Psi_1 
+ \frac{1}{2m}(D_+\Upsilon_2)^{\ast}D_+\Upsilon_2  \nn \\ 
&& \hspace*{1cm} + 2\lambda (\Phi_1\Psi_1^{\ast} -i \Phi_2^{\ast}\Upsilon_2)
(\Phi_1^{\ast}\Psi_1 + i \Phi_2\Upsilon_2^{\ast}) \biggr]\,. \label{H2}
\end{eqnarray}
This form of the Hamiltonian is manifestly semi-positive definite. In
relation to the $\mathcal{N}=2$ system, we note that by setting
$\Phi_1=\Upsilon_2=0$\,, the last term in (\ref{H2}) and the related
kinematic terms vanish and the Hamiltonian in \cite{LLM} can be
reproduced.

\medskip 

From the expression (\ref{H2}) it is easy to figure out the conditions
for the lowest energy solution. Those are given by the familiar ones
\begin{eqnarray}
D_1\Phi_1 = -i D_2\Phi_1\,, \quad D_1\Phi_2 = i D_2\Phi_2\,, \quad 
D_1\Psi_1 = i D_2\Psi_1\,, \quad D_1\Upsilon_2 = -i D_2\Upsilon_2\,, 
\end{eqnarray}
and the additional constraint 
\begin{eqnarray} 
\Phi_1\Psi_1^{\ast} = i\Phi_2^{\ast}\Upsilon_2\,. 
\end{eqnarray}
As a matter of course, the static soliton solution found in \cite{LLM}
satisfies these conditions when $\Phi_1=\Upsilon_2=0$\,. However, if we
try to turn on $\Phi_1$ and $\Upsilon_2$ non-trivially, then we cannot
fix $A_1$ and $A_2$ consistently to $\Phi_2$ and $\Psi_1$\,. It would be
interesting to look for solutions of different type. For example, spinor 
vortex solutions are disccused in \cite{spinor}.

\section{NR limit of $\mathcal{N}$=3 CSM - mixed cases} 

In the previous section we have discussed the case that all the
particles are kept and all the anti-particles are discarded. However
there is no reason why anti-particles are dropped off. Hence we shall
consider other NR limits containing anti-particles.

\medskip

Recall that the matter fields are expanded as in (\ref{expand}), 
\begin{eqnarray}
\phi_a &=&
\frac{1}{\sqrt{2m}}
\left[\,\e^{-im c^2 t} \Phi_a
+ \e^{im c^2 t} \hat{\Phi}^{\ast}_a \right] \qquad (a=1,2)\,, \nn \\
\psi &=& 
\sqrt{c}
\left[\,\e^{-i m c^2 t} \Psi
+ \e^{i m c^2 t} C \hat{\Psi}^{\ast} \right]\,, \nn \\
\chi &=& 
\sqrt{c}
\left[\,\e^{-i m c^2 t} \Upsilon
+ \e^{i m c^2 t} C \hat{\Upsilon}^{\ast}
\right]\,. \nn 
\end{eqnarray}
There are actually several choices to take the NR limits containing
anti-particles. We cannot, however, freely choose the matter content
kept in the NR limits if we take care of the consistency of the limits
with the parent theory. We will discuss the consistency of the NR limits
in section \ref{c}.

\medskip 

Here we shall pick up the following consistent cases:
\begin{enumerate}
\item the APPA case:
\[
\Phi_1=\hat{\Phi}_2=\hat{\Psi}=\Upsilon=0\,
\]
\item the PAPA case:
\[
\hat{\Phi}_1={\Phi}_2=\hat{\Psi}=\Upsilon=0\,.
\]
\end{enumerate}
The sequences of alphabets in the items indicate which of particle (P)
and anti-particle (A) is picked up in $(\phi_1,\phi_2,\psi,\chi)$\,,
respectively\footnote{In principle, we may keep both particle and
anti-particle in a single field (B) or neither of them (N). We will not
consider here these cases such as BAPN in this paper.}.

\medskip 

We will discuss each of the cases below. We will not touch the CS term
again and concentrate on the matter part only.

\subsection{A mixed case 1. - the APPA case }

Here let us consider the following mixed case:
\[
\Phi_1=\hat{\Phi}_2=\hat{\Psi}=\Upsilon=0\,.
\]
The matter content leads us to the following NR Lagrangian for the
matter fields
\begin{eqnarray}
\mathcal{L}_{\rm NR} &=& i\hat{\Phi}^{\ast}_1 \hat{D}_t\hat{\Phi}_1  
- \frac{1}{2m}(\hat{D}_i\hat{\Phi}_1)^{\ast} \hat{D}_i\hat{\Phi}_1 
+i\Phi^{\ast}_2 D_t \Phi_2 
- \frac{1}{2m}(D_i\Phi_2)^{\ast}D_i \Phi_2 \nn \\ 
&& +i \Psi_1^{\ast} D_t \Psi_1 
-\frac{1}{2m}(D_i\Psi_1)^{\ast} D_i\Psi_1 
+ i \hat{\Upsilon}_2^{\ast} \hat{D}_t \hat{\Upsilon}_2 
-\frac{1}{2m}(\hat{D}_i
\hat{\Upsilon}_2)^{\ast} \hat{D}_i\hat{\Upsilon}_2 \nn \\
&& -\frac{e}{2mc}F_{12}(|\Psi_1|^2  + |\hat{\Upsilon}_2|^2 )
- \lambda \left(|\hat{\Phi}_1|^4 - |\Phi_2|^4 \right)  \nn \\ 
&& + 3\lambda \left(|\hat{\Phi}_1|^2 + |\Phi_2|^2 \right)
\left(|\Psi_1|^2 - |\hat{\Upsilon}_2|^2 \right) \nn \\ 
&& - 4\lambda \left(|\hat{\Phi}_1|^2|\Psi_1|^2 
- |\Phi_2|^2|\hat{\Upsilon}_2|^2 \right) + \mathcal{O}(1/c^2)\,, 
\label{N=3-2}
\end{eqnarray}
where we have introduced another covariant derivative, which is friendly
to anti-particles,
\[
\hat{D}_{i} \equiv \partial_{i} - \frac{ie}{c}A_{i}\,, \qquad 
\hat{D}_t = c \hat{D}_0 = \partial_t -ie A_0\,. 
\]
In the derivation of (\ref{N=3-2}) we used the fermion equations of
motion
\begin{eqnarray}
\Psi_2 = -\frac{1}{2mc} D_+ \Psi_1\,, \qquad 
\hat{\Upsilon}_1= -\frac{1}{2mc} \hat{D}_- \hat{\Upsilon}_2\,
\end{eqnarray}
and removed $\Psi_2$ and $\hat{\Upsilon}_1$\,. Here we have also recombined $\hat{D}_i$ as 
\begin{eqnarray}
\hat{D}_{\pm} \equiv \hat{D}_1 \pm i\hat{D}_2\,. 
\end{eqnarray}
Comparing (\ref{N=3-2}) with (\ref{N=3-1}), we note that the signs of the 
charges of anti-particles are flipped and the last terms of (\ref{N=3-1}) are
missing. Again, one can reproduce the $\mathcal{N}$=2 action in
\cite{LLM} by setting $\hat{\Phi}_1=\hat{\Upsilon}_2=0$\,.

\subsubsection*{Schr\"odinger symmetry}

The NR Lagrangian with (\ref{N=3-2}) still has the Schr\"odinger
symmetry. The algebra can easily be derived in the same way as in the
previous section. For the bosonic generators the difference is just a
sign of the charge $e$ for anti-particles, so we will not repeat the
computation of the algebra here.

\medskip 

For the spin operators, the relative sign is not fixed since the
Lagrangian (\ref{N=3-2}) does not contain the term like the last term in
(\ref{N=3-1}). In the mixed case, however, it is possible to rotate
$\Psi_1$ and $\hat{\Upsilon}_2$ independently and there is an ambiguity
for the definition of their spins. Consequently, the undetermined
relative sign does not cause any problem.

\subsubsection*{$U(1)$ symmetries}

There are four $U(1)$ symmetries in this case and the number of each of
particles and anti-particles is conserved. The corresponding generators
are
\begin{eqnarray}
&& N_{\rm B1} = \int\!d^2x\,|\hat{\Phi}_1|^2\,, \quad 
 N_{\rm B2} = \int\!d^2x\,|\Phi_2|^2\,, \nn \\ 
&& N_{\rm F1} = \int\!d^2x\,|\Psi_1|^2\,, \quad 
 N_{\rm F2} = \int\!d^2x\,|\hat{\Upsilon}_2|^2\,. \nn 
\end{eqnarray}
Note that the mass operator $M$ is proportional to a sum of them and not
an independent quantity.

\subsubsection*{The positivity of the Hamiltonian}

The original Hamiltonian is given by 
\begin{eqnarray}
H &=& \int\!d^2x\,\biggl[
\frac{1}{2m}(\hat{D}_i\hat{\Phi}_1)^{\ast}\hat{D}_i\hat{\Phi}_1 
+ \frac{1}{2m}(D_i\Phi_2)^{\ast}D_i\Phi_2  \nn \\ 
&& \hspace*{1cm} + \frac{1}{2m}(D_i\Psi_1)^{\ast}D_i\Psi_1  
+ \frac{1}{2m}(\hat{D}_i\hat{\Upsilon}_2)^{\ast}
\hat{D}_i\hat{\Upsilon}_2 \nn \\ 
&& \hspace*{1cm} + \frac{e}{2mc}F_{12}(|\Psi_1|^2+|\hat{\Upsilon}_2|^2) 
+ \lambda(|\hat{\Phi}_1|^4-|\Phi_2|^4) \nn \\ 
&& \hspace*{1cm} -3\lambda (|\hat{\Phi}_1|^2+|\Phi_2|^2)
(|\Psi_1|^2 - |\hat{\Upsilon}_2|^2) 
\nn \\ && \hspace*{1cm} 
+ 4\lambda (|\hat{\Phi}_1|^2|\Psi_1|^2 
- |\Phi_2|^2|\hat{\Upsilon}_2|^2) \biggr]\,.
\end{eqnarray}
By using the Gauss law constraint 
\begin{eqnarray}
F_{12} = \frac{e}{\kappa}(-|\hat{\Phi}_1|^2 + |\Phi_2|^2 + |\Psi|^2 
- |\hat{\Upsilon}_2|^2)\,, 
\end{eqnarray}
this expression can be rewritten as 
\begin{eqnarray}
H &=& \int\!d^2x\,\biggl[
\frac{1}{2m}(\hat{D}_+\hat{\Phi}_1)^{\ast}\hat{D}_+\hat{\Phi}_1 
+ \frac{1}{2m}(D_-\Phi_2)^{\ast}D_-\Phi_2  \nn \\ 
&& \hspace*{1cm} + \frac{1}{2m}(D_-\Psi_1)^{\ast}D_-\Psi_1  
+ \frac{1}{2m}(\hat{D}_+\hat{\Upsilon}_2)^{\ast}\hat{D}_+\hat{\Upsilon}_2
\biggr]\,.
\end{eqnarray}
Thus, the Hamiltonian is semi-positive definite. The conditions for the
lowest energy solution are
\begin{eqnarray}
\hat{D}_+\hat{\Phi}_1 = D_-\Phi_2 = D_-\Psi_1 =
\hat{D}_+\hat{\Upsilon}_2 =0\,.
\end{eqnarray}

\subsubsection*{Supersymmetries}

The supersymmetry transformation at the leading order is given by
\begin{eqnarray}
&& \delta_1 \hat{\Phi}_1 = i\sqrt{2mc}\, \alpha_1^{(2)\ast}\hat{\Upsilon}_2\,, 
\qquad 
\delta_1 \Phi_2 =  -\sqrt{2mc}\, \alpha_1^{(1)\ast}\Psi_1\,,  
\nn \\
&& \delta_1 \Psi_1 = \sqrt{2mc}\, \alpha_1^{(1)}\Phi_2\,, 
\qquad 
\delta_1 \hat{\Upsilon}_2 = i\sqrt{2mc}\,\alpha_1^{(2)}\hat{\Phi}_1\,,  
\nn \\ 
&& \delta_1 A_0 = \frac{e}{\sqrt{2mc}\kappa} 
\Bigl[
\alpha_1^{(1)\ast}\Psi_1\Phi_2^{\ast} -\alpha_1^{(1)}\Psi_1^{\ast}\Phi_2
 -i\alpha_1^{(2)\ast}\hat{\Upsilon}_2\hat{\Phi}_1^{\ast} 
-i\alpha_1^{(2)}\hat{\Upsilon}_2^{\ast}\hat{\Phi}_1
\Bigr]\,, \nn \\  
&& \delta_1 A_i = 0\,. \nn  
\end{eqnarray}
The second supersymmetry transformation is
\begin{eqnarray}
&& \delta_2 \hat{\Phi}_1 = - \frac{i}{\sqrt{2mc}}\,
\alpha_1^{(1)\ast}\hat{D}_-\hat{\Upsilon}_2\,, 
\qquad 
\delta_2 \Phi_2 =
 -\frac{1}{\sqrt{2mc}}\,\alpha_1^{(2)\ast}D_+\Psi_1\,, 
\nn \\ 
&& \delta_2 \Psi_1 = -\frac{1}{\sqrt{2mc}}\,\alpha_1^{(2)}D_-\Phi_2\,, 
\qquad 
\delta_2 \hat{\Upsilon}_2 = \frac{i}{\sqrt{2mc}}\,\alpha_1^{(1)}
\hat{D}_+\hat{\Phi}_1\,,
\nn \\ 
&& \delta_2 A_0 = -\frac{e}{(2mc)^{3/2}\kappa} 
\Bigl[\alpha_1^{(2)\ast}(D_+\Psi_1)\Phi_2^{\ast} 
-\alpha_1^{(2)}(D_+ \Psi_1)^{\ast}\Phi_2 \nn \\ 
&& \hspace*{3.8cm} +i\alpha_1^{(1)\ast} \hat{D}_- 
\hat{\Upsilon}_2 \hat{\Phi}_1^{\ast} 
+i\alpha_1^{(1)}(\hat{D}_- \hat{\Upsilon}_2)^{\ast}\hat{\Phi}_1
\Bigr]\,,
\nn \\ 
&& \delta_2 A_+ = -\frac{2e}{\sqrt{2mc}\kappa}\left[
\alpha_1^{(1)\ast}\hat{\Upsilon}_2\hat{\Phi}_1^{\ast} 
-i\alpha_1^{(2)}\Psi_1^{\ast}\Phi_2
\right]\,, \nn \\ 
&& \delta_2 A_- = \frac{2e}{\sqrt{2mc}\kappa}\left[
\alpha_1^{(1)}\hat{\Upsilon}_2^{\ast}\hat{\Phi}_1 
+i \alpha_1^{(2)\ast}\Psi_1\Phi_2^{\ast} 
\right]\,. \nn  
\end{eqnarray} 

\medskip 

Now we should comment on the parameters of supersymmetry
transformation. Fist of all, $\alpha_2^{(1)}$ (or equivalently
$\alpha_2^{(2)}$) is not contained in the NR supersymmetry. Secondly,
$\alpha_1^{(1)}$ and $\alpha_1^{(2)}$ are not separated. That is, those
are common to both the leading and the next-to-leading
supersymmetries. According to the knowledge obtained in the previous
section, we may argue that all of the next-to-leading supersymmetries
are broken due to the interaction potential and all of the leading
supersymmetries are preserved. Indeed, that is the case. We can check
this statement explicitly by acting the supersymmetry transformations to
the Lagrangian with (\ref{N=3-2}). In addition, since there is no
next-to-leading supersymmetry, we have no superconformal symmetry. In
summary, we have two complex supercharges
\begin{eqnarray}
Q_1^{(1)}=\sqrt{2m}\int\! d^2x\, \Phi_2\Psi_1^*\,, \qquad 
Q_1^{(2)}=\sqrt{2m}\int\! d^2x\, \hat \Phi_1\hat\Upsilon_2^*\,, 
\end{eqnarray}
whose non-trivial Poisson brackets are given by 
\begin{eqnarray}
\{Q_1^{(1)}, Q_1^{(1)\ast}\} = -2mi (N_{\rm B2} + N_{\rm F1})\,, \cr
\{Q_1^{(2)}, Q_1^{(2)\ast}\} = -2mi (N_{\rm B1} + N_{\rm F2})\,.  
\end{eqnarray}

Thus, we have found a less supersymmetric \sch algebra in this case.
That is, the number of preserved supersymmetries is different according
to the choice of NR limits while the Schr\"odinger symmetry is still
preserved.

\subsection{A mixed case 2. - the PAPA case} 

As another possibility, let us consider the following mixed case:
\[
\hat{\Phi}_1={\Phi}_2=\hat{\Psi}=\Upsilon=0\,, 
\]
which leads us to the following NR Lagrangian for the matter fields
\begin{eqnarray}
\mathcal{L}_{\rm NR} &=& 
i{\Phi}^{\ast}_1 {D}_t{\Phi}_1  
-\frac{1}{2m}({D}_i{\Phi}_1)^{\ast} {D}_i{\Phi}_1 
+i\hat{\Phi}^{\ast}_2 \hat{D}_t\hat{\Phi}_2 
- \frac{1}{2m}(\hat{D}_i\hat{\Phi}_2)^{\ast}\hat{D}_i \hat{\Phi}_2 \nn \\ 
&& +i \Psi_1^{\ast} D_t\Psi_1 -\frac{1}{2m}(D_i\Psi_1)^{\ast} D_i\Psi_1 
+ i \hat{\Upsilon}_2^{\ast} \hat{D}_t\hat{\Upsilon}_2 
-\frac{1}{2m}(\hat{D}_i\hat{\Upsilon}_2)^{\ast}
\hat{D}_i\hat{\Upsilon}_2 \nn \\
&& 
-\frac{e}{2mc}F_{12}(|\Psi_1|^2  + |\hat{\Upsilon}_2|^2 )
- \lambda 
\left(|{\Phi}_1|^4 - |\hat{\Phi}_2|^4 \right)  \nn \\ 
&& + 3\lambda
\left(
|{\Phi}_1|^2 + |\hat{\Phi}_2|^2 
\right)
\left(|\Psi_1|^2 - |\hat{\Upsilon}_2|^2 \right) 
\nn \\ && 
 - 4\lambda \left(
|{\Phi}_1|^2|\Psi_1|^2 - |\hat{\Phi}_2|^2|\hat{\Upsilon}_2|^2 \right) \nn
\\ && +4\lambda (\Phi_1\hat{\Phi}_2 \Psi_1^* \hat{\Upsilon}_2^*
+ \Phi_1^* \hat{\Phi}_2^* \hat{\Upsilon}_2 \Psi_1)
+ \mathcal{O}(1/c^2)\,. \label{N=3-3}
\end{eqnarray}
In the derivation of (\ref{N=3-3}) we used the fermion equations of
motion
\begin{eqnarray}
\Psi_2 = -\frac{1}{2mc} D_+ \Psi_1\,, \qquad 
\hat{\Upsilon}_1= -\frac{1}{2mc} \hat{D}_- \hat{\Upsilon}_2\,
\end{eqnarray}
and removed $\Psi_2$ and $\hat{\Upsilon}_1$\,. 

\subsubsection*{Schr\"odinger symmetry}

The Lagrangian (\ref{N=3-3}) still has the Schr\"odinger symmetry. The
algebra can easily be derived in the same way as in the previous
section. For the bosonic generators the different point is just a sign
of the charge $e$ for anti-particles. To avoid the repetition, we will
not present the computation of the algebra here.

\subsubsection*{$U(1)$ symmetries}

There are three $U(1)$ symmetries. The corresponding generators are
\begin{eqnarray}
&& N_{\rm B} = \int\!d^2x\,\left(|{\Phi}_1|^2-|\hat{\Phi}_2|^2\right)\,,
\qquad 
 N_{\rm F} = \int\!d^2x\,\left(|\Psi_1|^2 -|\hat{\Upsilon}_2|^2\right)\,,\nn
\end{eqnarray}
and the mass operator $M$\,.

\subsubsection*{The positivity of the Hamiltonian}

The original Hamiltonian is given by 
\begin{eqnarray}
H &=& \int\!d^2x\,\biggl[
\frac{1}{2m}(D_i\Phi_1)^{\ast}D_i\Phi_1 
+ \frac{1}{2m}(\hat{D}_i\hat{\Phi}_2)^{\ast}\hat{D}_i\hat{\Phi}_2  \nn \\ 
&& \hspace*{1cm} + \frac{1}{2m}(D_i\Psi_1)^{\ast}D_i\Psi_1  
+ \frac{1}{2m}(\hat{D}_i\hat{\Upsilon}_2)^{\ast}
\hat{D}_i\hat{\Upsilon}_2 \nn \\ 
&& \hspace*{1cm} + \frac{e}{2mc}F_{12}(|\Psi_1|^2+|\hat{\Upsilon}_2|^2) 
+ \lambda(|\Phi_1|^4-|\hat{\Phi}_2|^4) \nn \\ 
&& \hspace*{1cm} -3\lambda (|\Phi_1|^2+|\hat{\Phi}_2|^2)
(|\Psi_1|^2 - |\hat{\Upsilon}_2|^2) 
\nn \\ && \hspace*{1cm} 
+ 4\lambda (|\Phi_1|^2|\Psi_1|^2 
- |\hat{\Phi}_2|^2|\hat{\Upsilon}_2|^2)  
-4\lambda (\Phi_1\hat{\Phi}_2\Psi_1^{\ast}\hat{\Upsilon}_2^{\ast}  
+ \Phi_1^{\ast}\hat{\Phi}_2^{\ast}\hat{\Upsilon}_2\Psi_1) \biggr]\,.
\end{eqnarray}
By using the Gauss law constraint 
\begin{eqnarray}
F_{12} = \frac{e}{\kappa}(|\Phi_1|^2 - |\hat{\Phi}_2|^2 + |\Psi|^2 
- |\hat{\Upsilon}_2|^2)\,, 
\end{eqnarray}
this expression can be rewritten as 
\begin{eqnarray}
H &=& \int\!d^2x\,\biggl[
\frac{1}{2m}(D_+\Phi_1)^{\ast} D_+ \Phi_1 
+ \frac{1}{2m}(\hat{D}_-\hat{\Phi}_2)^{\ast}\hat{D}_-\hat{\Phi}_2  \nn \\ 
&& \hspace*{1cm} + \frac{1}{2m}(D_+\Psi_1)^{\ast}D_+\Psi_1  
+
\frac{1}{2m}(\hat{D}_-\hat{\Upsilon}_2)^{\ast}\hat{D}_-\hat{\Upsilon}_2 
\nn \\ 
&& \hspace*{1cm} +4 \lambda(\Phi_1\hat{\Upsilon}_2^{\ast} 
+ \hat{\Phi}_2^{\ast}\Psi_1) 
(\Phi_1^{\ast}\hat{\Upsilon}_2 + \hat{\Phi}_2\Psi_1^{\ast})
\biggr]\,.
\end{eqnarray}
Thus, the Hamiltonian is semi-positive definite. The conditions for the
lowest energy solution are
\begin{eqnarray}
D_+\Phi_1 = \hat{D}_-\hat{\Phi}_2 = D_+\Psi_1 =
\hat{D}_-\hat{\Upsilon}_2 =0\,,
\end{eqnarray}
and the additional constraint
\begin{eqnarray}
\Phi_1 \hat{\Upsilon}_2^{\ast} = -\hat{\Phi}_2^{\ast}\Psi_1\,.
\end{eqnarray}

\subsubsection*{Supersymmetries}

The supersymmetry transformation at the leading order is given by
\begin{eqnarray}
&& \delta_1 \Phi_1 = \sqrt{2mc}\,\alpha_2^{(1)\ast} \Psi_1 \,, 
\qquad 
\delta_1 \hat{\Phi}_2 = -\sqrt{2mc}\,\alpha_2^{(1)} \hat{\Upsilon}_2 \,,  
\nn \\
&& \delta_1 \Psi_1 = -\sqrt{2mc}\,\alpha_2^{(1)}\Phi_1 \,, 
\qquad 
\delta_1 \hat{\Upsilon}_2 = \sqrt{2mc}\,\alpha_2^{(1)\ast}
\hat{\Phi}_2 \,,  
\nn \\ 
&& \delta_1 A_0 = -\frac{e}{\sqrt{2mc}\,\kappa} 
\Bigl[\alpha_2^{(1)\ast} \Psi_1 \Phi_1^* 
+ i\alpha_2^{(2)\ast}\Psi_1^* \Phi_1 
+ \alpha_2^{(1)\ast}\hat{\Upsilon}_2^* \hat{\Phi}_2 
+i \alpha_2^{(2)\ast} \hat{\Upsilon}_2 \hat{\Phi}_2^* 
\Bigr]\,, \nn \\  
&& \delta_1 A_i = 0\,. \nn  
\end{eqnarray}
and the one at the next-to-leading order is 
\begin{eqnarray}
&& \delta_2 \Phi_1 = \frac{1}{\sqrt{2mc}}\,\alpha_2^{(2)\ast}D_+\Psi_1\,, 
\qquad \delta_2 \hat{\Phi}_2 = \frac{1}{\sqrt{2mc}}\,
\alpha_2^{(2)}\hat{D}_-\hat{\Upsilon}_2\,, \nn \\ 
&& \delta_2 \Psi_1 = \frac{1}{\sqrt{2mc}}\,\alpha_2^{(2)}D_-\Phi_1  \,,
\qquad \delta_2 \hat{\Upsilon}_2 = \frac{1}{\sqrt{2mc}}\,\alpha_2^{(2)\ast}
\hat{D}_+\hat{\Phi}_2 \,,
\nn \\ 
&& \delta_2 A_0 = \frac{e}{(2mc)^{3/2}\kappa}\Bigl[
\alpha_2^{(2)\ast}D_+\Psi_1\Phi_1^{\ast} +
i\alpha_2^{(1)\ast}(D_+\Psi_1)^{\ast}\Phi_1 \nn \\ 
&& \hspace*{3.5cm} 
-\alpha_2^{(2)\ast}(\hat{D}_-\hat{\Upsilon}_2)^{\ast}\hat{\Phi}_2 
-i\alpha_2^{(1)\ast}\hat{D}_-\hat{\Upsilon}_2\hat{\Phi}_2^{\ast} 
\Bigr]\,,
\nn \\ 
&& \delta_2 A_+ =-\frac{2e}{\sqrt{2mc}\,\kappa}
(\alpha_2^{(1)\ast}\Psi_1^{\ast}\Phi_1 
+ \alpha_2^{(1)\ast}\hat{\Upsilon}_2\hat{\Phi}_2^{\ast}
)\,,
\nn \\ 
&& \delta_2 A_- = -\frac{2ie}{\sqrt{2mc}\,\kappa}
(\alpha_2^{(2)\ast}\Psi_1\Phi_1^{\ast} 
+ \alpha_2^{(2)\ast}\hat{\Upsilon}_2^{\ast}\hat{\Phi}_2)\,. 
\end{eqnarray}
First of all, note that the supersymmetry parameters
$\alpha_1^{(a)}~(a=1,2)$ are decoupled from the NR supersymmetry. Then,
$\alpha_2^{(1)}$ is not independent of $\alpha_2^{(2)}$ and so the
number of the independent supersymmetry parameters is 2 (in real). The
parameter is shared in both the leading and the next-to-leading
supersymmetries. Hence we guess that only the leading supersymmetry is
preserved and the next-to-leading is broken. As we can directly show, it
is really the case. Since the next-to-leading supersymmetry is broken,
the corresponding superconformal symmetry is also absent. In summary,
the theory we have considered here has an $\mathcal{N}$=1 supersymmetry
in 1+2 dimensions. The non-trivial supersymmetry algebra is given by
\begin{eqnarray}
\{Q_1, Q_1^*\} = -2iM\,,
\end{eqnarray}
where the supercharge is given by 
\begin{eqnarray}
Q_1&=&\sqrt{2m}\int\! d^2x\, 
\left[\Phi_1\Psi_1^* - \hat\Phi_2^* \hat\Upsilon_2\right]\,.
\end{eqnarray}

\subsection{Peculiar features of NR supersymmetry}

Summarizing the results obtained in this section, we may deduce the
following statements:
\begin{quotation}
\noindent {\it 1) A \sch symmetry always appears independently of the
choice of NR limits, but the number of preserved supersymmetries depends
on the choice. }
\end{quotation}
\begin{quotation}
\noindent {\it 2) The NR limit with only the particles (or
anti-particles) leads us to the same number of the supersymmetries as
the original relativistic theory. When including anti-particles, some of
the supersymmetries are broken or the supersymmetries are completely
broken.}
\end{quotation} 
\begin{quotation}
\noindent {\it 3) The supersymmetries in the original relativistic
theory are not enhanced after the NR limit.  }
\end{quotation}
\begin{quotation}
\noindent {\it 4) If supersymmetry parameters are not separated after
taking a NR limit, the leading supersymmetries are preserved but the
next-to-leading ones are broken.}
\end{quotation}
It would be interesting to check the observations for other models. 

\section{NR limit as consistent truncation}
\label{c}

Let us describe a consistency of NR limits in detail from the viewpoint
of the parent relativistic theory i.e. the $\mathcal{N}$=3 CSM
system. The discussion here confirms the consistency of the
non-relativistic limit taken in the previous sections.

\medskip 

In \cite{JP}, the authors addressed a consistency of NR limits in the
bosonic CSM system (Jackiw-Pi model). The original relativistic action
was expanded by using the field expansion like in (\ref{expand}), and
the conservation of the particle number and that of the anti-particle
number were checked. Since the anti-particle number is conserved
independently of the particle number, we can pick up a subsector with no
anti-particle.

\medskip 

The similar argument should be applied for supersymmetric CSM
systems. The $U(1)$ symmetries realized in the expanded action play an
important role in this argument. In the case of the $\mathcal{N}$=2 CSM
system \cite{LLM} there are four $U(1)$ symmetries and the number of
each of particles and anti-particles is conserved independently. Thus,
there is no problem to take any desired non-relativistic limit.
However, the $\mathcal{N}$=3 CSM system has a more complicated potential
and we should be careful about the $U(1)$ symmetries.

\subsubsection*{$U(1)$ symmetries and conserved quantities}

In order to check whether the conditions are satisfied or not, we have
to derive the expanded Lagrangian by substituting (\ref{expand}) into
(\ref{Lag}) without dropping off any of the particles and
anti-particles.

\medskip 

The expanded Lagrangian at the first order of $1/c$ is composed of the
kinetic terms for the bosons and fermions, $\mathcal{L}_{\rm B}$ and
$\mathcal{L}_{\rm F}$\,, respectively, and the Pauli interaction
$\mathcal{L}_{\rm Pauli}$\,, the four-boson interaction
$\mathcal{L}_{\rm 4B}$ and the boson-fermion interaction
$\mathcal{L}_{\rm BF}$ as follows:
\begin{eqnarray}
\CL&=&\CL_{\rm B} +\CL_{\rm F} +\CL_\mathrm{Pauli}+\CL_\mathrm{4B}
+\CL_\mathrm{BF}\,, \label{full} \\
\CL_{\rm B}&=&
i\Phi_1^* D_t \Phi_1
+i\hat\Phi_1^*\hat D_t \hat\Phi_1
-\frac{1}{2m}\left(
(D_i\Phi_1)^*D_i\Phi_1
+ (\hat{D}_i \hat \Phi_1)^*\hat D_i\hat \Phi_1
\right) \nn \\ 
&& + 
i\Phi_2^* D_t \Phi_2
+i\hat\Phi_2^*\hat D_t \hat\Phi_2
-\frac{1}{2m}\left(
(D_i\Phi_2)^* D_i\Phi_2 + 
(\hat{D}_i\hat \Phi_2)^* \hat D_i\hat \Phi_2\right)\,, \nn \\
\CL_{\rm F} &=& 
i\Psi_1^*D_t \Psi_1
+i\Psih_1^*\hat D_t \Psih_1
-\frac{1}{2m}\left(
(D_i\Psi)^*_1D_i\Psi_1
+(\hat{D}_i\Psih_1)^* \hat D_i\Psih_1
\right) \nn \\ 
&& + i\Upsilon_2^*D_t \Upsilon_2
+i\hat{\Upsilon}_2^*\hat D_t \hat{\Upsilon}_2
-\frac{1}{2m}\left(
(D_i\Upsilon_2)^* D_i\Upsilon_2 
+ (\hat{D}_i\hat{\Upsilon}_2)^* \hat D_i\hat{\Upsilon}_2
\right)\,, \nn \\ 
\CL_\mathrm{Pauli}&=&
-\frac{e}{2mc}F_{12}\left(
|\Psi_1|^2-|\Upsilon_2|^2
-|\Psih_1|^2+|\hat{\Upsilon}_2|^2
\right)\,, \nn  \\
\CL_\mathrm{4B} &=&
-\lambda\bigg[ 
\left(|\Phi_a|^2+|\Phih_a|^2\right)
\left(
|\Phi_1|^2-|\Phi_2|^2+|\Phih_1|^2-|\Phih_2|^2
\right) \nn  \\
&&~~~~
+2 (|\Phi_1|^2|\hat{\Phi}_1|^2-|\Phi_2|^2|\hat{\Phi}_2|^2) 
\bigg]\,, \nn 
\\
\CL_\mathrm{BF} &=& 3\lambda\left(
|\Phi_a|^2+|\Phih_a|^2
\right)
\left(
|\Psi_1|^2-|\Upsilon_2|^2
+|\Psih_1|^2-|\hat{\Upsilon}_2|^2
\right) \nn  \\
&&
+4\lambda\Bigg[
-|\Phi_1|^2|\Psi_1|^2
+\Phi_1\Phih_2\Psi_1^*\hat{\Upsilon}_2^*
-\Phi_1^*\Phih_2^*\Psi_1\hat{\Upsilon}_2
+|\Phih_2|^2|\hat{\Upsilon}_2|^2
\nn \\ 
&&~~~~
-|\Phih_1|^2|\Psih_1|^2
+\Phih_1^*\Phi_2^*\Psih_1\Upsilon_2
+\Phi_2\Phih_1\Upsilon_2^*\Psih_1^*
+|\Phi_2|^2|\Upsilon_2|^2
\nn \\&&~~~~
-|\Phi_1|^2|\Psih_1|^2
+|\Phi_2|^2|\hat{\Upsilon}_2|^2
-|\Phih_1|^2|\Psi_1|^2
+|\Phih_2|^2|\Upsilon_2|^2
\Bigg]
\nn \\ 
&&
-2i\lambda\Bigg[
2\Phi_1\Phih_1^*\Psi_1^*\Psih_1
+\Phi_1\Phi_2\Psi_1^*\Upsilon_2^*
+\Phih_1^*\Phih_2^*\Psih_1\hat{\Upsilon}_2
+2\Phi_2\Phih_2^*\Upsilon_2^*\hat{\Upsilon}_2
\nn \\ 
&&~~~~
+2\Phi_1^*\Phih_1\Psi_1\Psih_1^*
+\Phi_1^*\Phi_2^*\Psi_1\Upsilon_2
+\Phih_1\Phih_2\Psih_1^*\hat{\Upsilon}_2^*
+2\Phi_2^*\Phih_2\Upsilon_2\hat{\Upsilon}_2^*
\Bigg]\,. \label{last} 
\end{eqnarray}
Here we have used the fermion equations of motion and removed
$\Psi_2$\,, $\hat{\Psi}_2$\,, $\Upsilon_1$ and $\hat{\Upsilon}_1$\,.

\medskip 

From the Lagrangian (\ref{full}) we can figure out the $U(1)$
symmetries. The sensitive interactions are $+4\lambda [\ldots]$ and
$-2i\lambda[\ldots]$ in (\ref{last}). Without these interactions there
are eight $U(1)$ symmetries and the number of each field is
conserved. In particular, the $\mathcal{N}$=2 CSM system has no
problem. The interaction $+4\lambda [\ldots]$ yields eight $U(1)$
symmetries in a non-trivial way, but $-2i\lambda [\ldots]$ breaks half
of the $U(1)$'s. As a consequence, the four $U(1)$ symmetries are
preserved (Tab.\,\ref{u1s}).

\vspace*{0.5cm}
\begin{table}[htbp]
\begin{center}
\begin{tabular}{c|cccccccc}
 & $\Phi_1$ & $\hat{\Phi}_1$ & $\Phi_2$ & $\hat{\Phi}_2$ 
& $\Psi_1$ & $\hat{\Psi}_1$ & $\Upsilon_2$ & $\hat{\Upsilon}_2$ \\ 
\hline 
$U(1)_1$ & 1 & 0 & 0 & 0 & 1 & 0 & 0 & 0 \\ 
$U(1)_2$ & 0 & 0 & 1 & 0 & 0 & 0 & 1 & 0 \\ 
$U(1)_3$ & 0 & 0 & 0 & 1 & 0 & 0 & 0 & 1 \\
$U(1)_4$ & 0 & 1 & 0 & 0 & 0 & 1 & 0 & 0 
\end{tabular}
\caption{\footnotesize The remaining four $U(1)$ symmetries.} \label{u1s}
\end{center}
\end{table}

\medskip

From $U(1)_1$ and $U(1)_2$ the total number of $\Phi_1$\,, $\Phi_2$\,,
$\Psi_1$ and $\Upsilon_2$ is conserved. The total number of
$\hat{\Phi}_1$\,, $\hat{\Phi}_2$\,, $\hat{\Psi}_1$ and
$\hat{\Upsilon}_2$ is also conserved from $U(1)_3$ and $U(1)_4$.  Hence,
by setting the latter number to be zero, we can realize all particle
case (the PPPP case) consistently as in \cite{JP}. By using $U(1)_1$ and
$U(1)_3$\,, we understand that the PAPA case is also consistent to the
argument in \cite{JP}.

\medskip 

However, only from the viewpoint of the consistent truncation of the
matter content, one may allow a weaker condition. We propose the
following two criteria for the consistent truncation of the fields in NR
limits:
\begin{enumerate}
\item The total number of the matter fields picked up in the NR limit is
      conserved. \\ (strong condition)
\item The truncation of the field is consistent to all of the equations
      of motion. \\ (weak condition)
\end{enumerate}
Note that if the strong condition is satisfied then the weak condition
is also satisfied.

\medskip 

The PPPP case and the PAPA case satisfy the strong condition but the
APPA case does not. Still, the APPA case satisfies the weak
condition. Hence the truncation is consistent at the level of the
equations of motion and it would still make sense at least at the
classical level. There would possibly be a subtlety at quantum
mechanical level.

\medskip 

On the other hand, an exotic case such as the PPPA case is excluded even
by the weak condition. In addition, the supersymmetries are completely
broken.

\medskip 

Here, we have discussed the consistency of the matter field truncation
in NR limits. Nevertheless, it might be possible to consider a wider
class of NR limits if we take the stance to regard the NR limit as a
generation technique of super \sch invariant field theories. It would be
interesting to investigate whether the consistency of NR limits is
related to the consistency of the resulting NR theories.

\section{Summary and Discussion}

We have presented new super \sch invariant field theories by considering
NR limits of the $\mathcal{N}$=3 relativistic CSM system in 1+2
dimensions.

\medskip 

First, by taking a NR limit with only the particles, we have derived an
$\mathcal{N}$=3 super \sch invariant CSM system. By using the standard
Noether theorem, we explicitly constructed the generators of the super
\sch symmetry and computed the Poisson brackets of the generators.

\medskip 

Then, as other NR limits we have considered the two mixed cases: 1) the
APPA case and 2) the PAPA case. The bosonic \sch symmetry still persists
in both cases but the number of the preserved supersymmetries decreased
from $\mathcal{N}$=3 to $\mathcal{N}$=2 and $\mathcal{N}=1$\,
respectively. In particular, it is possible to find the supersymmetries
only at the leading, that is, no supersymmetries at the next-to-leading.

\medskip 

An interesting observation is that a less supersymmetric \sch algebra is
realized depending on the matter content held on in taking the NR limit.
In any NR limits the bosonic \sch symmetry is always preserved. On the
other hand, the number of the preserved supersymmetries depends on the
matter content in the NR limit. In particular, the inclusion of
anti-particles would be sensitive only to supersymmetries.

\medskip 

It would be interesting to study the other NR limits we have not
discussed here. In principle, it can be done and several super \sch
algebras would be obtained. As another direction, it would be
interesting to investigate the NR symmetry in a broken phase where
$\langle\phi\rangle \neq 0$\,, though we have discussed only in a
symmetric phase where $\langle\phi \rangle =0$\,.

\medskip

The full quantum treatment of the CSM theory should be investigated
further. In particular, the full \sch invariance might be broken by the
non-zero beta function for various coupling constants. With enough
supersymmetries ($\mathcal{N} =2$ or higher), we can argue that the beta
function for our models vanish to all orders in perturbation
theory. Thanks to the supersymmetry, all the coupling constants are
related to the charge $e^2/\kappa$ of the Chern-Simons theory, so we
only have to study the renormalization of the photon polarization
function. The perturbative corrections to the photon polarization
function, however, trivially vanish in the non-relativistic system just
because it does not allow particle/anti-particle pair creation, which
can easily be seen from the retarded nature of the Green functions for
matters. This guarantees the vanishing beta function for the
supersymmetric theories to all orders in perturbation theory. One can
also explicitly check that the perturbative beta function vanishes
\cite{Bergman:1993kq,Kim:1996rz}\footnote{One of the author Y.N. would
like to thank Y.~Nishida for discussions on this point.}. It would be
interesting to investigate further the situations in less supersymmetric
theories descended from the same supersymmetric parent theory.

\medskip 

The $\mathcal{N}$=3 NR CSM systems and their relatives, 
which contain a single gauge field, have been studied well here. 
The next is to study the CSM system containing two
gauge fields like \cite{GW,Sangmin,ABJM}. It would be nice to investigate 
NR limits of the ABJM theory \cite{ABJM}. 
We will report on this issue in the near future \cite{NSY}.

\medskip 

We hope that our method would be a clue to develop super \sch invariant
field theories and could formulate some basic techniques there.

\section*{Acknowledgements}

It is our pleasure to acknowledge helpful discussions with S.~Lee, Y.~Nishida
and A.~P.~Schnyder. The work of NY was supported in part by the National
Science Foundation under Grant No.\ PHY05-55662 and the UC Berkeley
Center for Theoretical Physics. The work of MS was supported in part by
the Grant-in-Aid for Scientific Research (19540324) from the Ministry of
Education, Science and Culture, Japan. The work of KY was supported in
part by the National Science Foundation under Grant No.\ PHY05-51164 and
JSPS Postdoctoral Fellowships for Research Abroad.

\section*{Appendix}

\appendix

\section{Dimensional analysis}

In order to take a NR limit, we need to recover the speed of light
$c$\,, so let us check dimensions of the fields and parameters contained
in a relativistic CSM system. For this purpose, it is enough to consider
the $\mathcal{N}$=2 CSM system \cite{LLW} since the terms appearing in
the $\mathcal{N}$=3 CSM system are almost the same as in the
$\mathcal{N}$=2 CSM system.

\medskip 

First of all, let us set $[\hbar]=1$\,. Since
\[
 [\hbar] = M L^2 T^{-1} =1\,,
\] 
we obtain 
\[
 [M]=TL^{-2}\,. 
\]
Accordingly, the action is dimensionless. 

\medskip 

From the kinematic term $\int dt d^2x \partial \phi \partial \phi$\,,
the dimension of the scalar field $\phi$ is
\[
 [\phi] = L^{-1/2}\,. 
\]
Note that the mass term $\int dt d^2x m^2 c^2 \phi^2$ is consistent.

\medskip 

Next let us consider the self-interaction terms of $\phi$\,. From the
quartic interaction $\int dt d^2x \frac{m e^2}{c\kappa}|\phi|^4$\,, we
can figure out
\[
 \left[\frac{e^2}{\kappa}\right] = L^2 T^{-2}\,. 
\]
Thus, the 6th order term $\int dt d^2x \frac{e^4}{c^4\kappa^2}|\phi|^6$
is consistent.

\medskip 

Now let us consider the dimension of the charge $e$ from the Coulomb
force $F$ in 1+2 dimensions:
\[
 F=ma = -\frac{e^2}{r}\,.
\]
It is easy to see that 
\[
 [e^2]=T^{-1} \qquad \mbox{and} \qquad [\kappa]= T L^{-2}\,. 
\]
Note that the CS coupling is dimensionful. 

\medskip 

Then, from the covariant derivative $D_{\mu} = \partial_{\mu} +
i\frac{e}{c}A_{\mu}$\,, we find
\[
 [A_{\mu}]= T^{-1/2}\,. 
\]       
Thus, the CS terms are also consistent\footnote{In (2.1a) of \cite{LLM},
$\kappa/4c$ should be replaced by $\kappa/4$\,. This replacement is
consistent to (2.8) of \cite{LLM}\,.}.  For the fermionic field, the
argument is the similar.

\subsubsection*{Non-relativistic case} 

After taking a NR limit, it is natural to set $[M]=1$ and then
\[
 [T]=[L^2]\,.
\] 
This is appropriate to a NR field theory with dynamical exponent $z=2$
such as a \sch invariant field theory\,. Note that the CS coupling
$\kappa$ becomes dimensionless.

\section{The detail of the spinor rotation}

In the original paper \cite{N=3} the gamma matrices are given in the
Majorana representation
\[
 \gamma_{\rm M}^0 = -i\sigma_2\,, \quad \gamma_{\rm M}^1=\sigma_3\,, 
\quad \gamma_{\rm M}^2 = \sigma_1\,,
\]
while each of the two fermion fields is a 2-component complex fermion
defined as
\[
\left(\begin{array}{c}
\psi \\ 
\chi
\end{array}
\right)
= \psi_1 + i\psi_2\,, 
\]
where $\psi_k~(k=1,2)$ are real 4-component Majorana fermions i.e.,
$(\psi_k)^{\ast}=\psi_k$\,.

\medskip 

In order to take a non-relativistic limit, it is convenient to move from
the Majorana representation to the Dirac one (\ref{Dirac})\,. This can
be done by using the following transformation:
\[
 \gamma^{\mu} =U^{-1}\gamma^{\mu}_{\rm M} U\,, \qquad U=U_1 U_2\,, \quad 
U_1 = \frac{1}{\sqrt{2}}
(1+i\sigma_1)\,, \quad U_2 = \frac{1}{\sqrt{2}}(1+i\sigma_3)\,.
\]
The following relations are available:
\[
U^TU=i\sigma_1\,, \qquad U^{-1}U^{\ast}=-i\sigma_1\,.
\]

\medskip 

Associated with this rotation, the original complex fermions $\psi_{\rm
o}$ and $\chi_{\rm o}$ are also rotated to a new set of the complex
fermions $\psi$ and $\chi$\, in the Lagrangian.

\subsubsection*{The condition for $\alpha_2$ - the rotated Majorana condition}

We should be careful for the rotation in rewriting the supersymmetry
transformation because a part of the supersymmetry parameters is given
in terms of Majorana spinor in the original paper \cite{N=3}.

\medskip 

To see the connection between their Majorana condition and our
condition, we note that our spinor $\alpha_2$ is related to
$\alpha_{o2}$ as
\[
 \alpha_2 = \left(\begin{array}{c}
\alpha_2^{(1)} \\ \alpha_2^{(2)}
\end{array}
\right)
= U^{-1}\alpha_{o2} = U^{-1}\left(\begin{array}{c}
\alpha_{o2}^{(1)} \\ \alpha_{o2}^{(2)}
\end{array}
\right)\,. 
\] 
By defining new real spinors
\[
 \alpha_o^{(\pm)} \equiv \frac{1}{2}\left(\alpha_{o2}^{(1)} \pm 
\alpha_{o2}^{(2)}\right)\,,
\]
we obtain
\[
 \alpha_2^{(1)} = \alpha_o^{(-)} -i\alpha_o^{(+)}\,, \qquad 
 \alpha_2^{(2)} = \alpha_o^{(+)} -i\alpha_o^{(-)}\,. 
\]
Thus, in our notation, the Majorana condition reads
\begin{eqnarray}
\alpha_2^{(2)} =-i(\alpha_2^{(1)})^{\ast}\,, 
\end{eqnarray}
and $\alpha_2^{(1)}$ and $\alpha_2^{(2)}$ are not independent. As an
immediate consequence, the number of independent components of
$\alpha_2$ is 2 in real.

\end{document}